\documentclass[aps,pre,twocolumn,superscriptaddress,noshowpacs,longbibliography]{revtex4-2}
\usepackage{amsmath}
\usepackage{amssymb}
\usepackage{tikz}
\usepackage{graphicx}
\usepackage{dcolumn}
\usepackage{bm}
\usepackage{epsfig}
\usepackage{psfrag}
\usepackage{orcidlink}

\newcommand{\reffig}[1]{\mbox{Fig.~\ref{#1}}}
\newcommand{\refeq}[1]{\mbox{Eq.~(\ref{#1})}}
\newcommand{\refsec}[1]{\mbox{Sec.~\ref{#1}}}

\newcommand{\be}{\begin{equation}}
\newcommand{\ee}{\end{equation}}
\newcommand{\bal}{\begin{align}}
\newcommand{\eal}{\end{align}}
\newcommand{\bea}{\begin{eqnarray}}
\newcommand{\eea}{\end{eqnarray}}

\newcommand{\mm}{\mathrm{mm}}
\newcommand{\cm}{\mathrm{cm}}

\begin{document}

\title{\bf Experimental study of coupled quantum billiards with integrable and chaotic classical dynamics and test of a special Rosenzweig-Porter model}
\author{Xiaodong Zhang\,\orcidlink{0000-0002-9173-7109}}
\email{xiaodongzhang2021@gmail.com}
\affiliation{Lanzhou Center for Theoretical Physics, Key Laboratory of Theoretical Physics of Gansu Province, and Key Laboratory of Quantum Theory and Applications of MoE, Lanzhou University, Lanzhou, Gansu 730000, China}
\affiliation{Center for Theoretical Physics of Complex Systems, Institute for Basic Science, Daejeon 34126, Republic of Korea}
\affiliation{Shandong Key Laboratory of Space Environment and Exploration Technology, College of Physics and Electronic Engineering, Qilu Normal University, Jinan 250200, China}
\author{Jiongning Che\,\orcidlink{0000-0001-8512-8991}}
\affiliation{Yangtze Delta Region Institute, University of Electronic Science and Technology of China, Huzhou 313001, China}
\affiliation{Lanzhou Center for Theoretical Physics, Key Laboratory of Theoretical Physics of Gansu Province, and Key Laboratory of Quantum Theory and Applications of MoE, Lanzhou University, Lanzhou, Gansu 730000, China}
\author{Barbara Dietz\,\orcidlink{0000-0002-8251-6531}}
\email{bdietzp@gmail.com}
\affiliation{Center for Theoretical Physics of Complex Systems, Institute for Basic Science, Daejeon 34126, Republic of Korea}
\affiliation{Basic Science Program, Korea University of Science and Technology (UST), Daejeon 34113, Korea}
\affiliation{Lanzhou Center for Theoretical Physics, Key Laboratory of Theoretical Physics of Gansu Province, and Key Laboratory of Quantum Theory and Applications of MoE, Lanzhou University, Lanzhou, Gansu 730000, China}

\date{\today}
\bigskip
\begin{abstract}
	We report on the experimental study of the spectral properties of quantum systems consisting of two quantum billiards ({\bf QB}s), one with chaotic, the other one with integrable classical dynamics, that are coupled to each other via an opening in a common wall. They are compared to those of a special case of the Rosenzweig-Porter model with random matrices composed of two diagonal blocks modeling the spectral properties of the {\bf QB}s, that are coupled with a tunable parameter. We demonstrate that this model is suitable for the description of the experimental data and thus may be employed to determine the strength of the coupling. It results from the increasing overlap of eigenmodes in the {\bf QB}s penetrating through the opening into the other one, leading to a mixing of their eigenstates, and the breaking of the symmetry present in the {\bf QB} with integrable dynamics. This implicates deviations of the spectral properties from those of typical quantum systems with integrable and chaotic dynamics, respectively, and approaches those of a fully chaotic system for sufficiently large coupling strength. In contrast in previous studies the transition from integrable to chaotic dynamics was induced by introducing a random potential of increasing strength into such a {\bf QB} and applicability of a variant of the Rosenzweig-Porter model was tested.          
\end{abstract}

\maketitle 
\section{Introduction\label{Intro}}
We investigated experimentally the spectral properties of quantum systems consisting of two parts, one with integrable, one with chaotic classical dynamics, that are coupled to each other. The experiments are performed with a flat microwave resonator emulating a quantum billiard ({\bf QB}), a paradigm model for the study of aspects of quantum chaos, because its classical dynamics only depends on its shape~\cite{Sinai1970,Bunimovich1979,Berry1981,LesHouches1989,Haake2018}. 

The objective of quantum chaos for single-particle quantum systems with a well-defined classical limit is to obtain information on the classical dynamics exclusively from the spectral properties and the properties of the wave functions of the corresponding quantum system. According to the Berry-Tabor conjecture~\cite{Berry1977a}, the eigenvalues of typical integrable systems exhibit Poisson statistics. On the other hand, the renowned Bohigas-Giannoni-Schmit conjecture~\cite{Bohigas1984} (see also~\cite{Casati1980,Berry1981a}) states that the spectral properties of typical quantum systems, whose classical analogue is chaotic, are well described by random-matrix theory ({\bf RMT}), namely those of Hermitian random matrices from one of the Wigner-Dyson ensembles associated with Dyson's threefold way~\cite{Dyson1962,Mehta2004}. These are the Gaussian Orthogonal Ensemble (GOE) if time-reversal invariance is preserved, the Gaussian unitary ensemble (GUE) if it is violated, and the Gaussian Symplectic Ensemble (GSE) for quantum systems with symplectic symmetry. A similar quantum-chaotic behavior was observed in quantum systems with no well-defined classical limit, like relativistic neutrino billiards~\cite{Berry1987}. This and deviations from the behavior expected for the corresponding nonrelativistic quantum billiard can be understood based on a semiclassical approach~\cite{Dietz2020,Yupei2022,Zhang2021}.     

Wigner introduced {\bf RMT} to describe the properties of eigenstates in complex many-body quantum systems and suggested a connection between their spectral properties and those of random matrices~\cite{Wigner1951,Wigner1955,Wigner1957,Porter1965,Brody1981,Haq1982,Guhr1989,Weidenmueller2009}. In the past four decades {\bf RMT} has been successfully employed for the analysis of aspects of single-particle and many-body quantum chaos~\cite{LesHouches1989,Guhr1998,StoeckmannBuch2000,Haake2018,Gomez2011,Frisch2014,Mur2015,Altland2024}. Furthermore, the Wigner-Dyson ensembles have been generalized to {\bf RMT} ensembles, referred to as Rosenzweig-Porter model, describing systems with mixed regular-chaotic dynamics~\cite{Rosenzweig1960,Pandey1981,Pandey1995,Guhr1996,Kunz1998,Pino2019,Khaymovich2020,Berkovits2020,Zhang2023b} or to transition ensembles interpolating between two universality classes~\cite{Pandey1981,Pandey1991,Altland1993,Dietz2009,Dietz2019b}, and extended to the tenfold way~\cite{Altland1997} and beyond~\cite{Kawabata2019}. 

We use flat, cylindrical microwave resonators for the experimental study of universal fluctuation properties in the eigenvalue spectra of {\bf QB}s. For microwave frequencies below the cutoff frequency $f^{\rm max}=c/2h$, where $h$ is the height of the resonator and $c$ is the speed of light, only transverse magnetic modes are excited inside the resonator, and the electrical-field vector is perpendicular to the top and the bottom plates of the resonator. Then, the Helmholtz equation describing the electromagnetic field inside the resonator is scalar and mathematically identical to the two-dimensional Schr\"odinger equation for the {\bf QB} of corresponding shape with Dirichlet boundary conditions along the contour~\cite{Sinai1970,Bunimovich1979,Berry1981,LesHouches1989,Haake2018}. Accordingly, in that frequency range the microwave resonator is referred to as microwave billiard ({\bf MB})~\cite{Stoeckmann1990,Sridhar1991,Graef1992,Deus1995,StoeckmannBuch2000}. To obtain complete sequences of eigenvalues, the experiments are performed at low temperature ($\lesssim 5$~K) with superconducting {\bf MB}s. 

In recent works we realized experimentally quantum systems that undergo a transition from Poisson to GUE statistics~\cite{Zhang2023b}, also via GOE statistics~\cite{Zhang2024}. There, the transition was realized by introducing a random potential into the resonator that induces complexity and partial time-reversal invariance violation. The spectral properties were compared to those of the Rosenzweig-Porter model~\cite{Rosenzweig1960}. In the present work we study situations where the spectral properties of a quantum system with chaotic classical dynamics are affected by contributions from a system with integrable classical dynamics or vice versa the effect of contributions of a chaotic system to a system with integrable dynamics. For this we employ two {\bf QB}s that are coupled via an opening of varying size in a common part of their walls. This billiard system can be mapped onto a {\bf RMT} model which is a special case of the Rosenzweig-Porter model. In~\cite{Dietz2006a} we employed this model to investigate symmetry breaking in coupled quantum-chaotic billiards. In the present experiments the {\bf QB}s are of similar size, and they are coupled with varying strength to analyze spectral properties for very weak to moderate coupling. We find that long-range correlations are already affected for extremely weak coupling, similarly to those of quantum systems comprising quantum-scarred modes~\cite{McDonald1979,Heller1984,LesHouches1989}. Furthermore, the {\bf RMT} model is suitable for the determination of the strength of the coupling. 

The experimental setup is presented in~\refsec{Experiment} and in~\refsec{RMT} the {\bf RMT} model which is constructed from it. Then, in~\refsec{SpectralProp} we present experimental and {\bf RMT} results for the spectral properties. Finally, in~\refsec{Concl} we summarize and discuss the results.

\section{Experimental setup\label{Experiment}}
\begin{figure}[!htbp]
\centering
\includegraphics[width=0.47\linewidth]{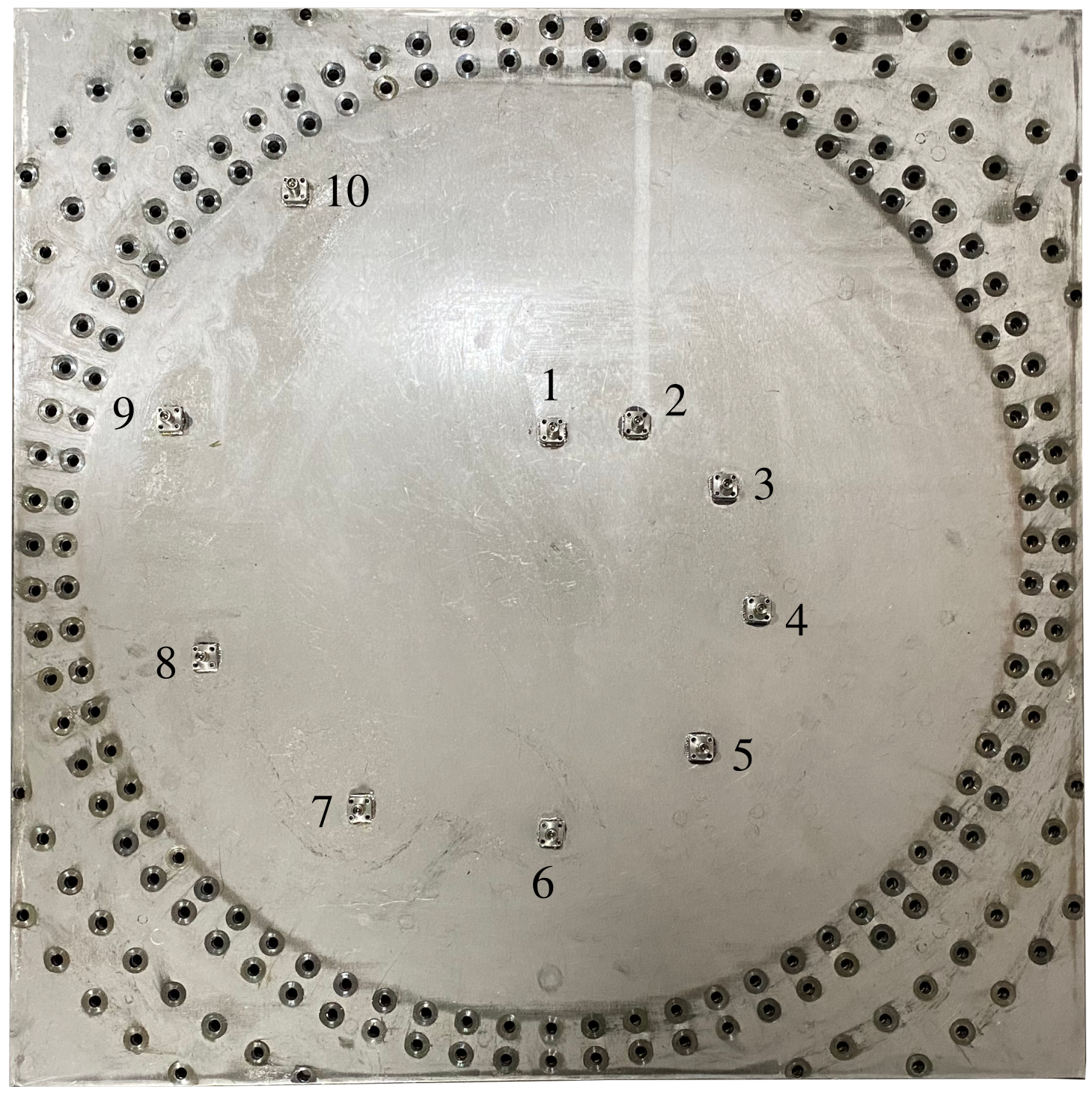} 
\includegraphics[width=0.47\linewidth]{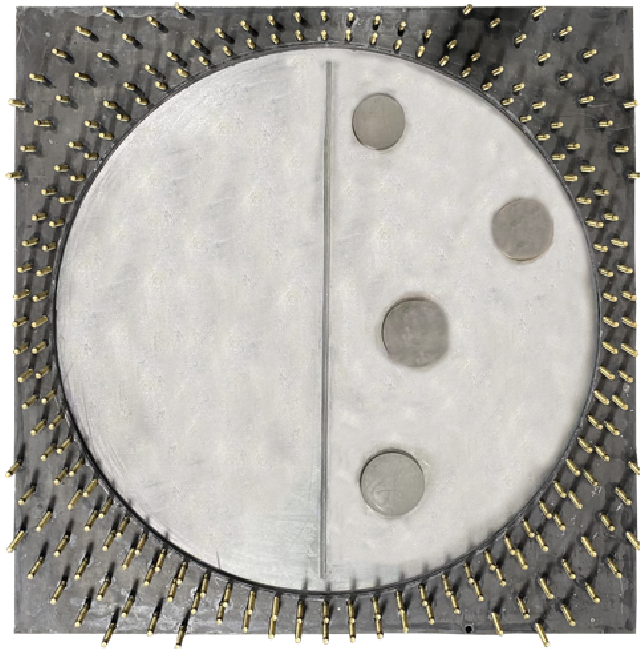}
\caption{Left: Photograph of the niobium top plate with 10 antenna ports. Right: Photograph of the basin, i.e. the cavity without lid, consisting of a niobium bottom plate and a brass plate with a circular hole, into which a niobium rod is inserted along the diameter, dividing it into a semicircular part and a part which contains four circular niobium scatterers. }  
\label{Photo_Exp}
\end{figure}

The experiments were performed at superconducting conditions with the same circular {\bf MB} as in Refs.~\cite{Zhang2023b,Zhang2024} with modification of components inserted into it. Figure~\ref{Photo_Exp} shows photographs of the lid and basin of the {\bf MB}. The resonator is composed of three 5~mm thick plates. Details are provided in the appendix in~\refsec{Details}. The top and bottom plate are made from niobium, whereas the middle one is a brass plate, which has a circular hole of radius $R=250$~mm forming the resonator body whose sidewall is coated with lead, and grooves in its top and bottom surface along that hole. Furthermore, screw holes were milled out of all plates along circles. The resonator height $h=5$~mm corresponds to a cutoff frequency $f^{\rm max}=30$~GHz. The cavity was separated along the diameter by a wall constructed from rectangular niobium rods of varying size into a semicircular part and a nearly-semicircular one. The heights and widths of the rods equal that of the cavity, 5~mm, and their lengths equal 500~mm, 476~mm or 470~mm, respectively. The rods were positioned in the cavity such that they leave at their ends close to the circular wall openings of width 0~mm, 12~mm and 15~mm, respectively. The three ${\bf MB}s$ are referred to as {\bf MB1}, {\bf MB2}, and {\bf MB3}, respectively, in the following. The classical dynamics of a semicircle billiard is integrable. The dynamics of the nearly semicircular part is non-integrable~\cite{Bunimovich1979,Ree1999}. To attain a billiard with chaotic dynamics, we inserted into that part four circular niobium disks with radii of 24, 27, 29, and 30~mm, respectively~\cite{Zhang2019,Zhang2023b}, corresponding to circular holes in the corresponding billiard. Niobium and lead become superconducting at $T_c=9.2$~K and $T_c=7.2$~K, respectively. To attain superconductivity with a high quality factor of $Q\gtrsim 10^4$, tin-lead was filled into the grooves for a good electrical contact, the three plates were screwed together tightly through the screw holes (cf. left and right part of~\reffig{Photo_Exp}), and the cavity was cooled down to below $\approx 5$~K in a cryogenic chamber constructed by ULVAC Cryogenics in Kyoto, Japan. 

The resonance frequencies $f_n, f_1\leq f_2\leq\dots$ of the {\bf MB}s yielding the eigenvalues of the corresponding {\bf QB}, are given by the positions of the resonances in the reflection and transmission spectra. For their measurement ten antenna ports were fixed to the lid as visible in the left part of~\reffig{Photo_Exp} and antennas were attached to them. These were connected via a switch to a Keysight N5227A Vector Network Analyzer (VNA) with SUCOFLEX126EA/11PC35/1PC35 coaxial cables~\cite{Zhang2023b}. Here the lengths of the antennas, that reached 0.5~mm into the cavity, were chosen as short as possible to ensure smallest possible openness of the cavity~\cite{Dietz2015a,StoeckmannBuch2000,Dembowski2005}. Note that for too long antennas the circular cavity emulates a singular billiard with intermediate statistics~\cite{Dietz2022,Zhang2024}. That absorption into the walls and affects from the antennas are negligible was tested for example in~\cite{Zhang2023b}, where the resonance frequencies were determined in superconducting experiments with the empty, circular microwave cavity and spectral properties were compared to those of the analytically determined eigenvalues. We found excellent agreement in the relevant frequency range.
 The VNA emits microwaves into the resonator via one antenna $a$ and receives them at the same antenna or another one, $b$, and computes their relative amplitude $\vert S_{ba}\vert$ and phase $\phi_{ba}$. Accordingly, it provides the complex scattering ($S$)-matrix elements, $S_{ba}=\vert S_{ba}\vert e^{i\phi_{ba}}$. 

Measurement at superconducting conditions was crucial to attain isolated resonances in the considered frequency ranges. These are well described by the complex Breit-Wigner form, $S_{ba}=\delta_{ba}-i\sqrt{\gamma_{na}\gamma_{nb}}/(f-f_n+(i/2)\Gamma_n)$. Here, $\gamma_{na}$ and $\gamma_{nb}$ denote the partial widths associated with antennas $a$ and $b$ and $\Gamma_n$ is the total resonance width, which is small since absorption into the walls of the {\bf MB}, scatterers and rod are negligible in superconducting cavities~\cite{Dietz2015a}. Determination of all resonance parameters is only possible after calibration of the $S$ matrix~\cite{Dembowski2005}, which under superconducting conditions requires a special cumbersome procedure~\cite{Marks1991,Rytting2001,Yeh2013,Dietz2024}. However, calibrated $S$-matrix elements are not needed for the determination of the resonance frequencies $f_n$. To identify complete sequences of resonance frequencies we fitted the Breit-Wigner form to the resonances for all antenna positions. Note that resonances may not be identified for too strongly overlapping resonances, which are absent in the present experiments. Above all, resonances are missing or are barely excited when the electric-field strength is too small at the position of an antenna. Accordingly, to cover all regions of the {\bf MB}, we chose the antenna positions along a spiral from close to the circle center to the circle wall.  
	
\begin{figure}[!htbp]
\centering
\includegraphics[width=1.0\linewidth]{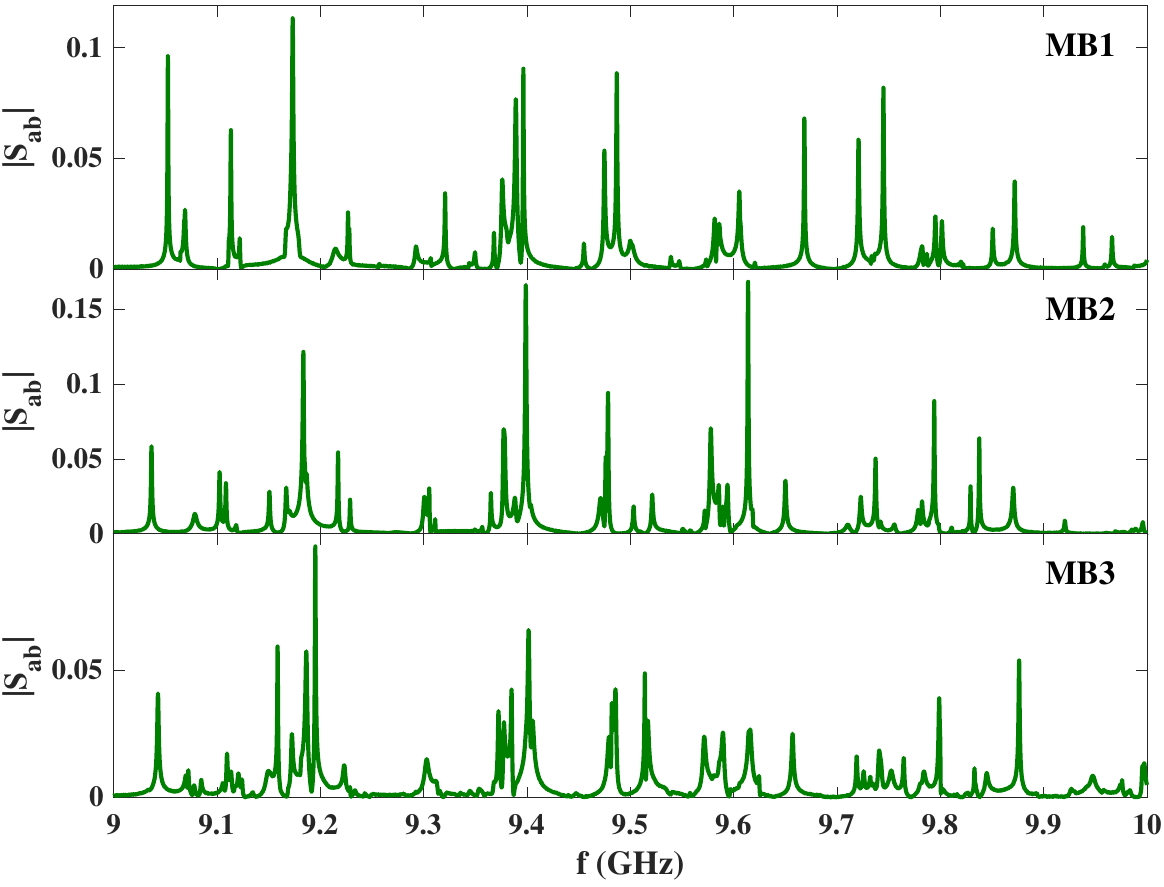} 
\caption{Amplitudes of the transmission matrix element $S_{ba}(f)$ as function of the microwave frequency $f$ from one of the 10 attached antennas $b$ to another one, $a$, in the frequency range from 9 to 10 GHz. From top to bottom the resonance spectra for rod lengths 500~mm, 476~mm and 470~mm, corresponding to weak, moderate and strong coupling of the modes excited in the two parts of the cavity are shown. Shown are the spectra for the transmission from antenna 7 to antenna 10 within the semicircular part.}  
\label{Resonance_Spectra}
\end{figure}

\begin{figure}[!htbp]
\centering
\includegraphics[width=1.0\linewidth]{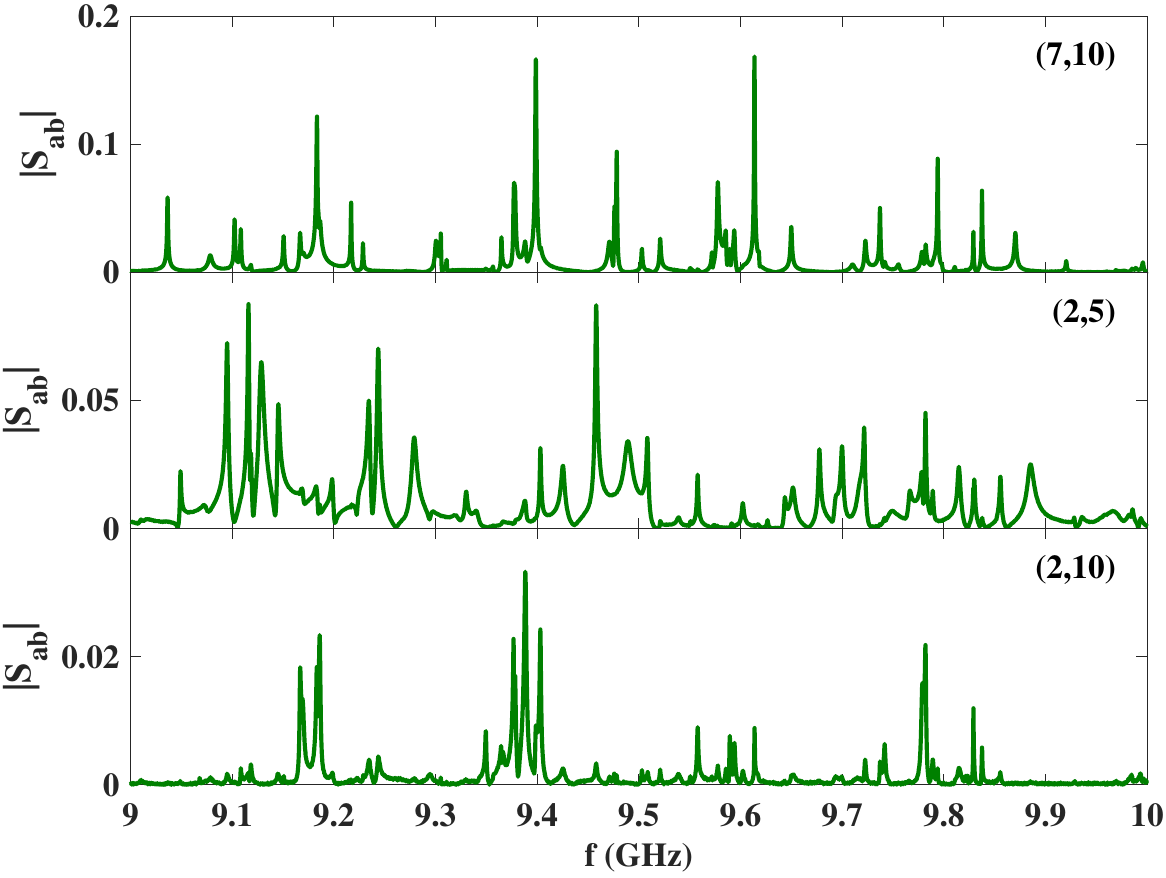}
	\caption{Same as~\reffig{Resonance_Spectra} for rod length 476~mm, corresponding to moderate coupling of the modes excited in the two parts of the cavity are shown. The top panel shows the transmission from antenna 7 to antenna 10 within the empty semicircular part, the middle panel that for transmission from antenna 5 to antenna 2 within the part with scatterers, and the bottom panel for transmission from antenna 2 to 10 between the two parts.}
\label{Resonance_Spectra_MB2}
\end{figure}

In~\reffig{Resonance_Spectra} we compare measured transmission spectra $|S_{ab}|$ of, from top to bottom, the {\bf MB1}, {\bf MB2} and {\bf MB3}. These were measured with antennas 7 and 10 which are both located in the semicircular part; cf.~\reffig{Photo_Exp} and~\reffig{Model} in the appendix. A significant shift in the peak positions is observed from top to bottom, i.e., when increasing the opening width, corresponding to a notable alteration in the resonance frequencies. Indeed, the opening in the wall allows the penetration of the electric-field modes excited in one half of the circular cavity into the other one, and thus a coupling to the modes excited there, which increases with increasing opening. Note that even for a rod length equal to the diameter of the circular cavity there is a nonzero, yet weak coupling between the resonator modes of the two parts of the cavity. The reason is that the rods did not fit perfectly flush with the top and bottom plates, implying that there is microwave leakage from one part of the cavity to the other one, even when there is no opening. The total leakage is much smaller for no opening, i.e. for {\bf MB1} compared to those with an opening, referred to as moderate and strong coupling for the {\bf MB2} and {\bf MB3}, respectively, in the following. In~\reffig{Resonance_Spectra_MB2} we compare for the case of moderate coupling, {\bf MB2}, transmission spectra obtained for antennas 7 and 10 in the semicircular parts with those measured between antennas 2 and 5 in the part containing the scatterers and antennas 2 and 10 for transmission between both parts. For the latter case the resonance peaks might be significantly smaller than in the other two spectra. Thus, to identify all resonance frequencies we used all measured spectra.   

In~\cite{Dembowski2001a,Dietz2009a} similar constructions were used to investigate at room temperature exceptional points and the effect of the rod and size of the slit on the scattering properties of an open microwave cavity with the shape of a quarter-circle, respectively. The coupling between the two parts of the microwave billiard is induced by the overlap of their eigenmodes, which is nonvanishing only if the field intensities are nonvanishing in both parts. Therefore, in order to ensure that this is the case for almost all eigenmodes, we chose the slits close to the circular boundary. To intuitively illustrate the effect of the opening, i.e., the implications of the coupling of modes excited in the two parts of the cavity, we computed for various resonance frequencies with COMSOL Multiphysics electric-field distributions, i.e., wave functions of the corresponding {\bf QB}.
\begin{figure}[!htbp]
\centering
\includegraphics[width=0.85\linewidth]{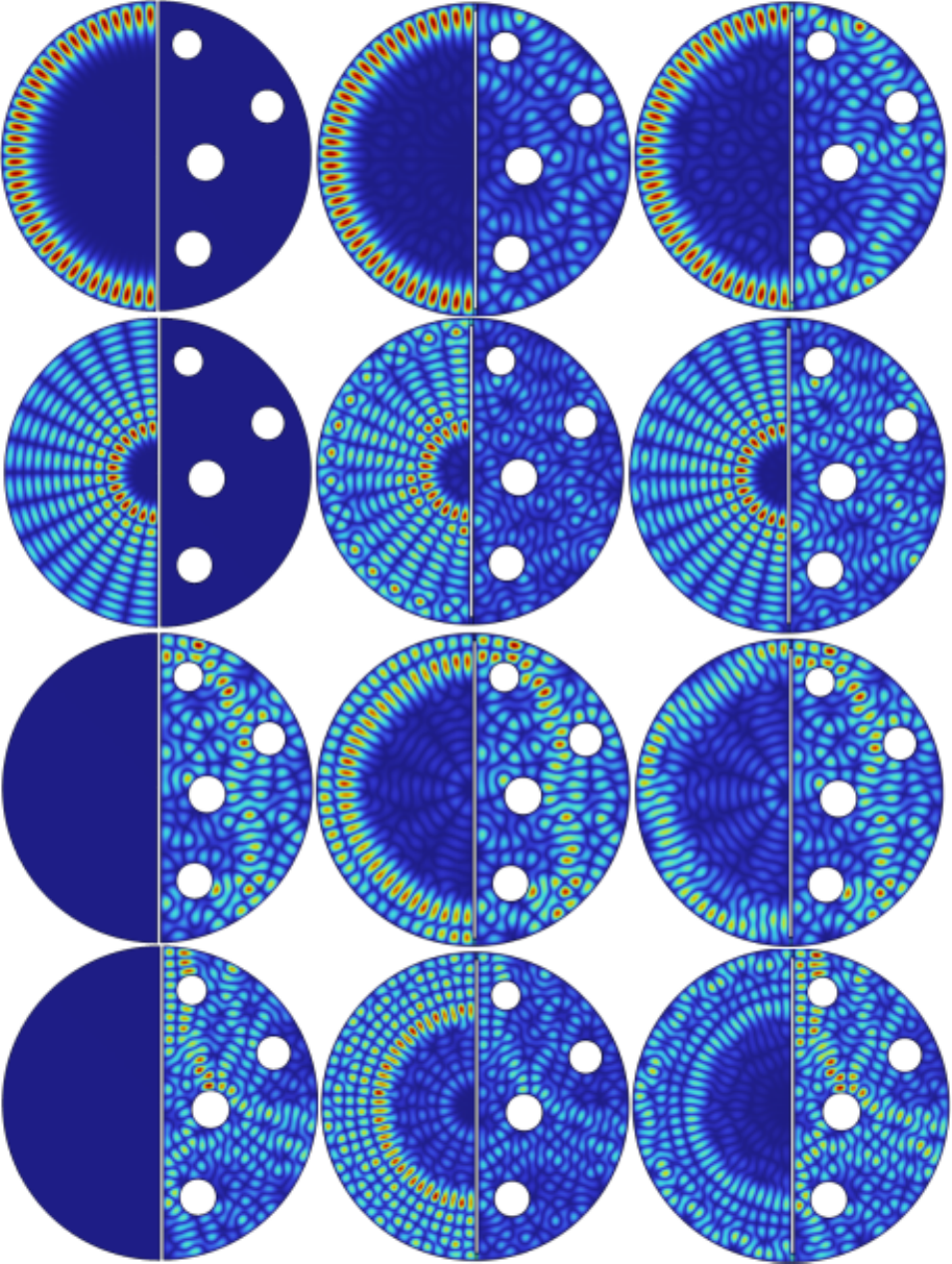}  
\includegraphics[width=0.05\linewidth]{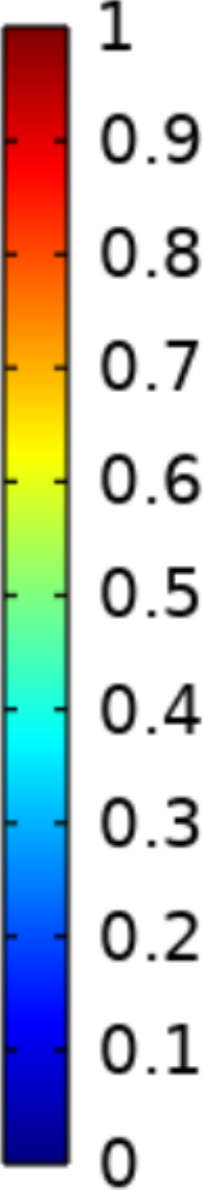}  
\caption{Intensity distributions of wave functions for the three considered sizes of the opening in the wall separating the cavity into a semicircular apart and a part, which contains four circular niobium disks. The intensity distributions were divided by their maximum value, so that they take values between 0 and 1. The first, second and third column correspond to the cases of weak, moderate and strong coupling, respectively. The resonance frequencies $f_n$ correspond to the same eigenmode number $n$ for the three configurations. They equal 8.5018 GHz, 8.4991 GHz, and 8.4943 GHz in the first row, 10.2945 GHz, 10.2940 GHz, and 10.2946 GHz in the second one, 9.0961 GHz, 9.0955 GHz, and 9.0954 GHz in the third one, and 11.0833 GHz, 11.0794 GHz, and 11.0818 GHz in the fourth row.}  
\label{Wavefunctions}
\end{figure}
In~\reffig{Wavefunctions} are shown intensity distributions of the electric-field modes for weak, moderate and strong coupling in the left, middle, and right columns, respectively for four different eigenmode numbers $n$. The corresponding resonance frequencies $f_n$ are given in the caption. For $f\gtrsim 8.5$~GHz the intensity distributions in {\bf MB2} and {\bf MB3} clearly reveal the presence of coupling and its impact on the wave patterns. For the {\bf MB1} eigenmodes are confined to one part of the cavity, since in the COMSOL Multiphysics calculations the contact between the wall and the top and bottom plates is assumed to be perfect, implying that the two parts are not coupled. For the other two cases the electric-field distributions penetrate with increasing coupling from one part into the other one and induce an increasing complexity of the intensity patterns in the semicircular part. This indicates that the properties of the eigenmodes are significantly altered when inducing the coupling. We demonstrate in~\refsec{SpectralProp} that the coupling affects the spectral properties. They are compared to those of a special case of the Rosenzweig-Porter model~\cite{Rosenzweig1960} which is introduced in the next section.  

\section{The {\bf RMT} model\label{RMT}}
The spectral properties of the {\bf MB}s are emulated with random matrices of the form  
\be
H=\begin{pmatrix}H_1&0\\0&H_2\end{pmatrix}+\alpha\begin{pmatrix}0&V\\V^T&0\end{pmatrix},
\label{RMTModel}
\ee
which is a variant of the Rosenzweig-Porter model introduced in~\cite{Rosenzweig1960}. The spectral properties of $H_1$, which has dimension $N_1$ and is diagonal with Gaussian-distributed random numbers of zero mean as entries, coincide with those of Poisson random numbers~\cite{Cadez2024}, that is, with those of typical quantum systems with integrable classical dynamics. Thus, it models the spectral properties of the empty semicircular cavity. The variance of the Gaussian distribution is chosen such that the matrix elements take values from the interval $[-1,1]$. The matrix $H_2$ is a random matrix with dimension $N_2$ drawn from the GOE, emulating the spectral properties of a typical, classically chaotic quantum system with preserved time-reversal invariance, that is, those of the part containing scatterers. Like the elements of $H_1$ those of $H_2$ are Gaussian distributed with zero mean. The variance is chosen such, that its eigenvalues take values from the same range $[-1,1]$ as the diagonal elements of $H_1$, $\left\langle {H_2}_{ij}^2 \right\rangle = \frac{1}{4N_2}(1 + \delta_{ij})$ with $\delta_{ij}$ denoting the Kronecker delta. The coupling matrix $V$ is a $N_1 \times N_2$ matrix, whose elements are generated by the same Gaussian distribution as the off-diagonal matrix elements of $H_2$. It couples the matrix elements of $H_1$ to those of $H_2$, that is, emulates the penetration of modes from one part of the {\bf MB} to the other one, leading to a coupling of the two systems. The strength of the coupling is tuned through the parameter $\alpha$. To obtain a dimension-independent parameter $\lambda$, $\alpha$ is rescaled such that for large $N_1\simeq N_2$ $H$ becomes a GOE matrix,  
\begin{equation}
\alpha = \frac{2}{\sqrt{N_2}}\lambda.
\label{eq2}
\end{equation}
We demonstrate in~\refsec{SpectralProp} that the model~\refeq{RMTModel} describes the spectral properties of the {\bf MB}s and thus provides a tool to estimate the strength of the coupling of the eigenmodes excited in the two parts of the cavity in terms of $\lambda$. Actually, in the experiments $N_1$ and $N_2$, which correspond to the number of eigenmodes in these two parts in the considered frequency range, $f\in [8,12]$~GHz, are similar, as explained in~\refsec{SpectralProp}. In~\cite{Dietz2014} a similar model was employed~\cite{Bohigas1993,Leyvraz1996,Schlagheck2006,Pecora2011} for the analysis of the spectral properties of a constant-width billiard exhibiting dynamical tunneling. However, in that case the number of modes of the integrable part was considerably smaller than that of the chaotic one, and also their coupling was significantly smaller.    

\section{Spectral properties of the {\bf MB}s\label{SpectralProp}}
Before comparing the spectral properties of the resonance frequencies of the {\bf MB}s with {\bf RMT} predictions for universal properties of typical quantum systems, system-specific properties need to be eliminated, that is, the resonance frequencies need to be unfolded such that the mean spacing is unity, i.e., the spectral density is uniform in the considered frequency range. This is achieved by replacing the resonance frequencies by the smooth part of the integrated spectral density, $\epsilon_n=\mathcal{N}^{\rm smooth}(f_n)$, which gives the number of resonance frequencies $n$ below $f=f_n$. Below the cutoff frequency, $\mathcal{N}^{\rm smooth}(f)$ is given by the Weyl formula~\cite{Weyl1912}, 
\be
\mathcal{N}^{\rm smooth}(f)=\frac{\mathcal{A}\pi}{c^2}f^2-\frac{\mathcal{L}}{2c}f+N_0,
\ee
with $\mathcal{A}$ and $\mathcal{L}$ denoting the area and perimeter of the billiard, respectively. In~\refsec{Details} of the appendix the areas of the three billiard geometries are provided. To determine $\mathcal{N}^{\rm smooth}(f_n)$ we fit a quadratic polynomial to the number of resonance frequencies $\mathcal{N}(f_n)$ below $f_n,n=1,\dots N_{\rm max}$ with $N_{\rm max}$ denoting the number of identified resonance frequencies. Here, we chose for the coefficient of the quadratic term the value of $\mathcal{A}$ allowing a variation of $3\%$ and the other coefficients as fit parameters. This yielded $\frac{\pi}{c^2}(\mathcal{A}^{\rm fit}-\mathcal{A})(f_{\rm max}^2-f_{\rm 1}^2)=0.003,1.8,0.442$ for the cases of weak, moderate and strong coupling, respectively, implying, that if at all, only one resonance frequency is missing in the second case. Furthermore, to check whether resonance frequencies are missing, we analyzed the fluctuating part of the spectral density, $\mathcal{N}^{\rm fluc}(f_n)=\mathcal{N}(f_n)-\mathcal{N}^{\rm smooth}(f_n)$. Corresponding curves are exhibited in~\reffig{Nfluc} in the apendix. They clearly fluctuate around zero and do not exhibit unusual jumps resulting from missing or spurious resonance frequencies.

We analyzed short-range correlations in the resonance-frequency spectra in terms of the distribution $P(s)$ of nearest-neighbor spacings $s_i=\epsilon_{i+1}-\epsilon_i$ and the integrated distribution $I(s)=\int_0^sds^\prime P(s^\prime)$, which has the advantage that it does not depend on the binning size of the histograms yielding $P(s)$. Furthermore, we analyzed the distribution of the ratios~\cite{Oganesyan2007,Atas2013,Atas2013a} of consecutive spacings between next-nearest neighbors, $r_j=\frac{\epsilon_{j+1}-\epsilon_{j}}{\epsilon_{j}-\epsilon_{j-1}}$. In~\cite{Oganesyan2007} the distributions of the ratios $\tilde r_j=\min\left(r_j,\frac{1}{r_j}\right)$, which take values between zero and unity, were considered. Ratios are dimensionless so that unfolding is not required. Furthermore, we evaluated the variance 
\be
\Sigma^2(L)=\left\langle\left[\mathcal{N}(L)-\langle\mathcal{N}(L)\rangle\right]^2\right\rangle
\ee 
of the number of unfolded eigenvalues in an interval of length $L$, and the rigidity 
\be
\Delta_3(L)=\left\langle\min_{a,b}\int_{\epsilon -L/2}^{\epsilon +L/2}d\epsilon\left[N(\epsilon )-a-b\epsilon\right]^2\right\rangle
\ee
as measures for long-range correlations~\cite{Mehta2004}. Here, $\langle\cdot\rangle$ denotes the spectral average over different parts of an eigenvalue spectrum. Other characteristics of long-range correlations are the spectral form factor~\cite{Mehta2004,Altland2025}, given by 
\be
K(t)= \frac{1}{N_{\rm max}}\left\langle\left\lvert N_{\rm max}^{-1}\small\sum_{n}^{N_{\rm max}}e^{2\pi i \epsilon_{n}t}\right\rvert^2\right\rangle.\label{eq:SFF}
\ee
and the power spectrum, 
\be
\label{PowerS}
s(\tau)=\left\langle\left\vert\frac{1}{\sqrt{N_{\rm max}}}\sum_{q=0}^{N_{\rm max}-1} \delta_q\exp\left(-2\pi i\tau q\right)\right\vert^2\right\rangle ,
\ee
for a sequence of $N_{\rm max}$ levels. Here, $0\leq\tau\leq 1$ and $\delta_q=\epsilon_{q+1}-\epsilon_1-q$ denotes the deviation of the $q$th nearest-neighbor spacing from its mean value $q$. The power spectrum exhibits for $\tau\ll 1$ a power-law dependence $\langle s(\tau)\rangle\propto \tau^{-\gamma}$~\cite{Relano2002,Faleiro2004,Riser2017,Riser2020}.  For integrable systems $\gamma =2$ and for chaotic ones $\gamma =1$ independently of the universality class of the quantum system~\cite{Gomez2005,Salasnich2005,Santhanam2005,Relano2008,Faleiro2006,Mur2015,Cadez2024,Bialous2016,Che2021}. 

The long-range measures are related to the two-point cluster function $Y_2(r)=1-R_2(r)$, where $R_2(r)$ is the two-point correlation function which is obtained from the experimental and numerical eigenvalues $\epsilon_n$ via the relation~\cite{Mehta2004,Dietz2022}
\be
R_2 (r)=\sum_{n=0}^\infty P(n,r)\ .
\ee
Here, $P(n,r)$ denotes the $(n+1)$st nearest-neighbor spacing distribution~\cite{Bogomolny1999,Bohigas1999}. The number variance $\Sigma^2(L)$ and the rigidity $\Delta_3(L)$ are related to the two-point cluster function as   
\be
\Sigma^2(L)=L-2\int_0^L(L-r)Y_2(r)dr\, ,
\ee
\be
\Delta_3(L)=\frac{L}{15}-\frac{1}{15L^4}\int_0^L(L-r)^3\left(2L^2-9rL-3r^2\right)Y_2(r)dr,
\ee
and the spectral form factor is given by $K(t)=1-b(t)$, where $b(t)$ is the Fourier transform of $Y_2(r)$~\cite{Mehta2004},
\be
b(t)=\int_{-\infty}^\infty Y_2(r)e^{-irt}dr.
\ee
\begin{figure}[!htbp]
\centering
\includegraphics[width=0.9\linewidth]{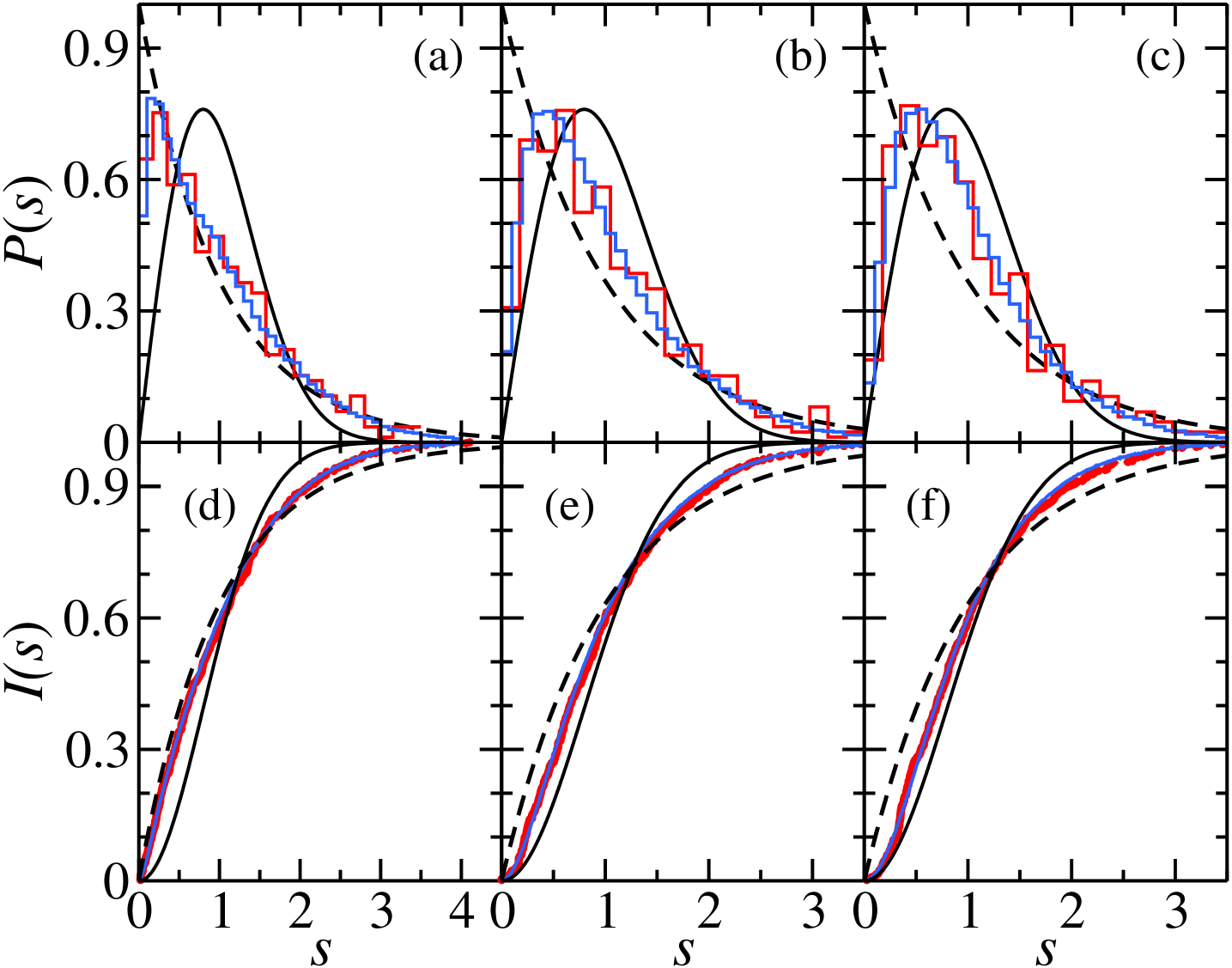}  
	\caption{Nearest-neighbor spacing distribution $P(s)$ [(a)-(c)] and the corresponding cumulative distribution $I(s)$ [(d)-(f)] for weak [(a),(d)], moderate [(b),(e)] and strong [(c),(f)] coupling. The red histograms and dots were obtained from the experimental data, the blue ones from the {\bf RMT} model~\refeq{RMTModel} for coupling strengths (a) $\lambda =0.03$, (b) $\lambda =0.23$, and (c) $\lambda=0.35$, respectively. These are compared with corresponding distributions for Poisson random numbers (black dashed lines) and random matrices from the GOE (black solid lines).}  
\label{NND}
\end{figure}
\begin{figure}[!htbp]
\centering
\includegraphics[width=0.9\linewidth]{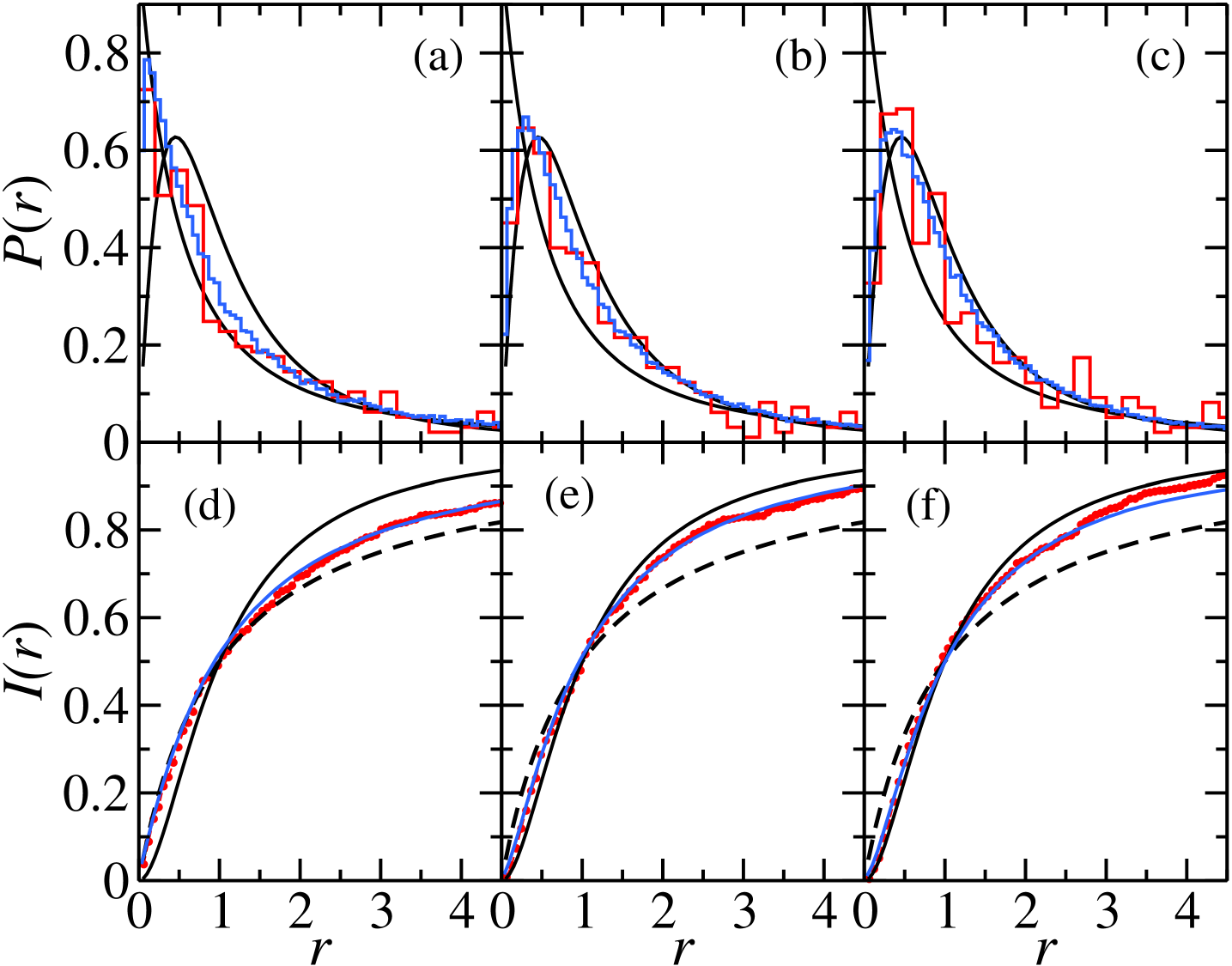}  
\caption{ Same as~\reffig{NND} for the ratio distribution $P(r)$.}
\label{Ratios}
\end{figure}
\begin{figure}[!htbp]
\centering
\includegraphics[width=0.9\linewidth]{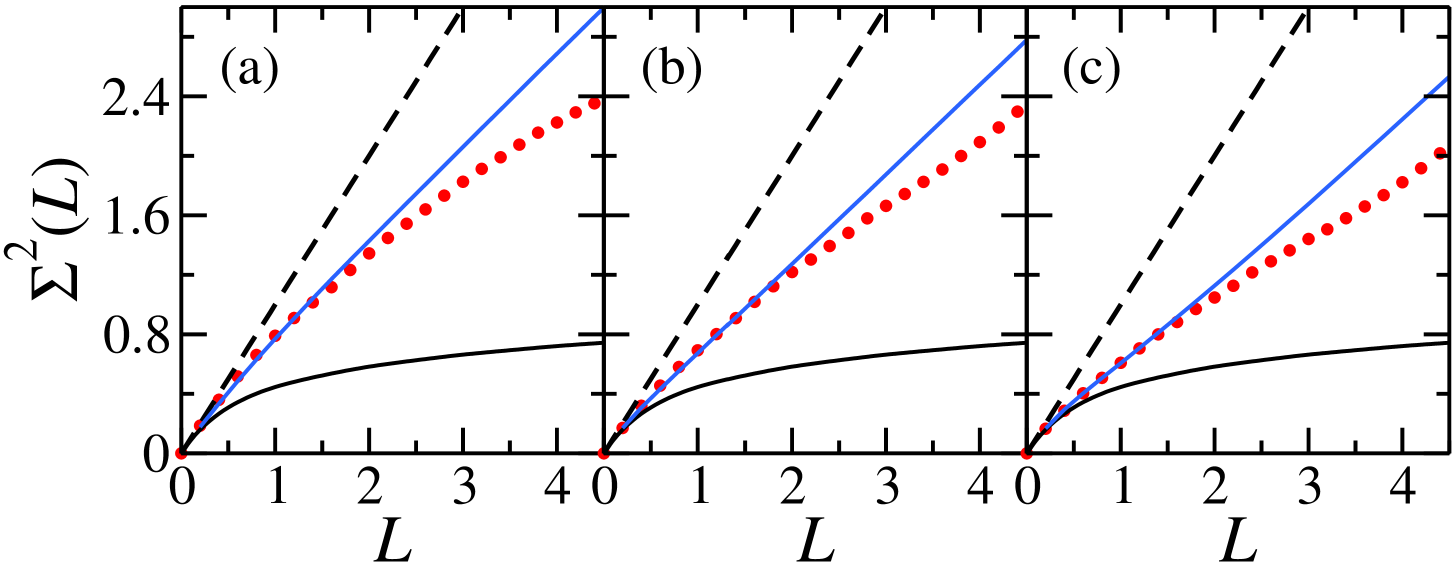}  
\caption{Same as~\reffig{NND} for the number variance $\Sigma^2(L)$.}
\label{Sigma2}
\end{figure}
\begin{figure}[!htbp]
\centering
\includegraphics[width=0.9\linewidth]{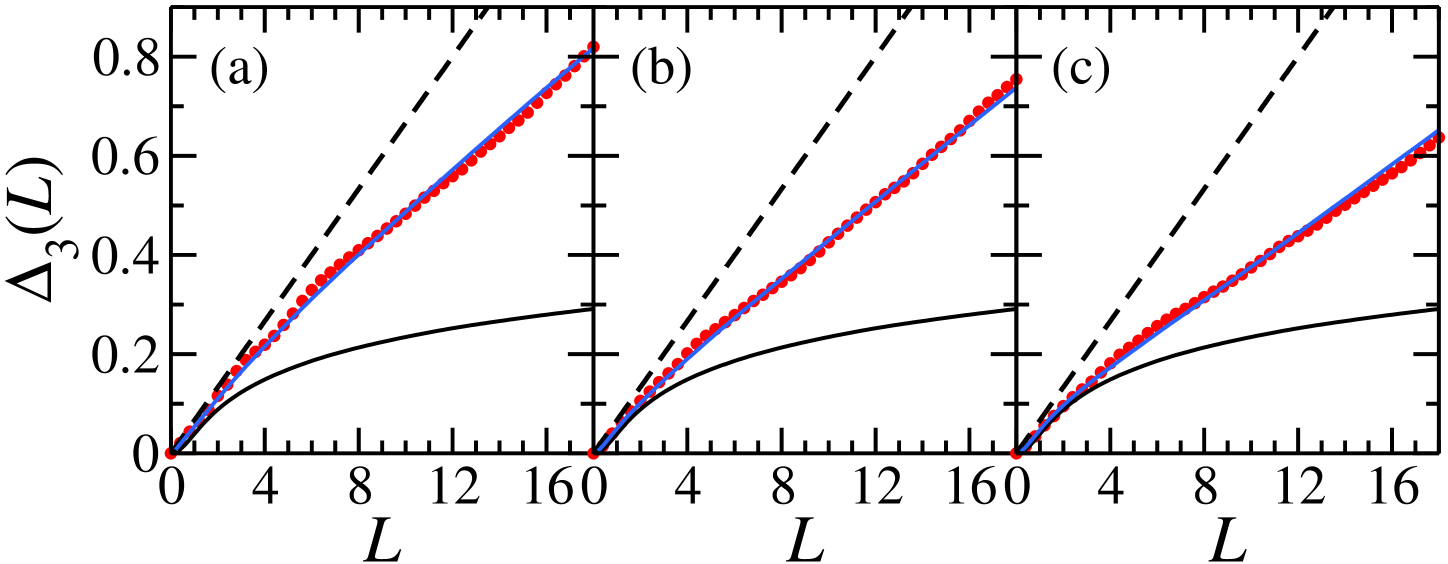} 
\caption{Same as~\reffig{NND} for the rigidity $\Delta_3(L)$.}
\label{Delta3}
\end{figure}
\begin{figure}[!htbp]
\centering
\includegraphics[width=0.9\linewidth]{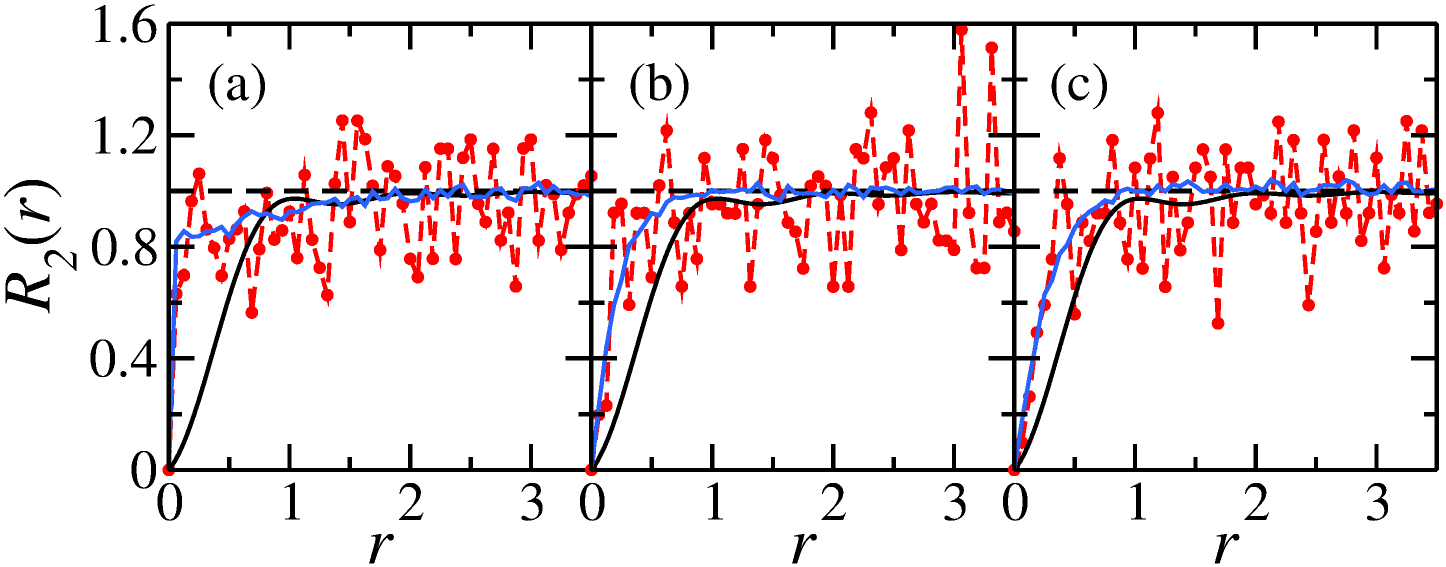} 
\caption{Same as~\reffig{NND} for the two-point correlation function $R_2(r)$.}
\label{R2}
\end{figure}
\begin{figure}[!htbp]
\centering
\includegraphics[width=0.9\linewidth]{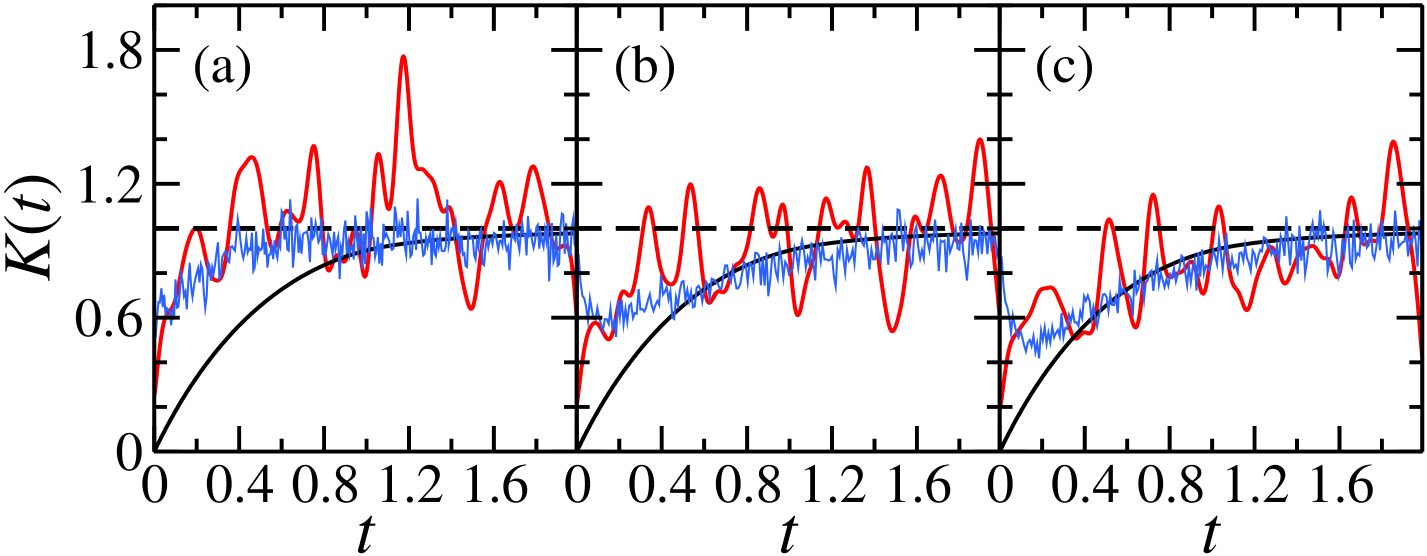}
\caption{Same as~\reffig{NND} for the spectral form factor $K(t)$.} 
\label{SFF}
\end{figure}
\begin{figure}[!htbp]
\centering
\includegraphics[width=0.9\linewidth]{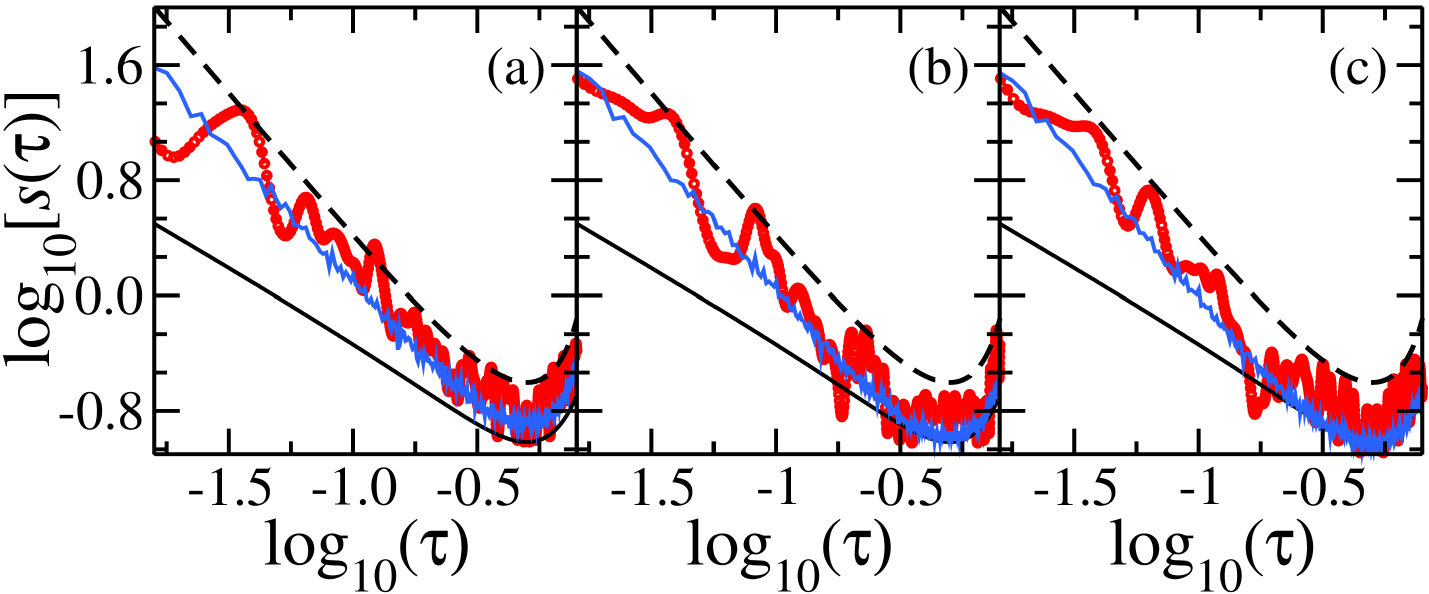}
\caption{Same as~\reffig{NND} for the power spectrum $s(\tau)$.}
\label{Power}
\end{figure}
\begin{figure}[!htbp]
\centering
\includegraphics[width=0.9\linewidth]{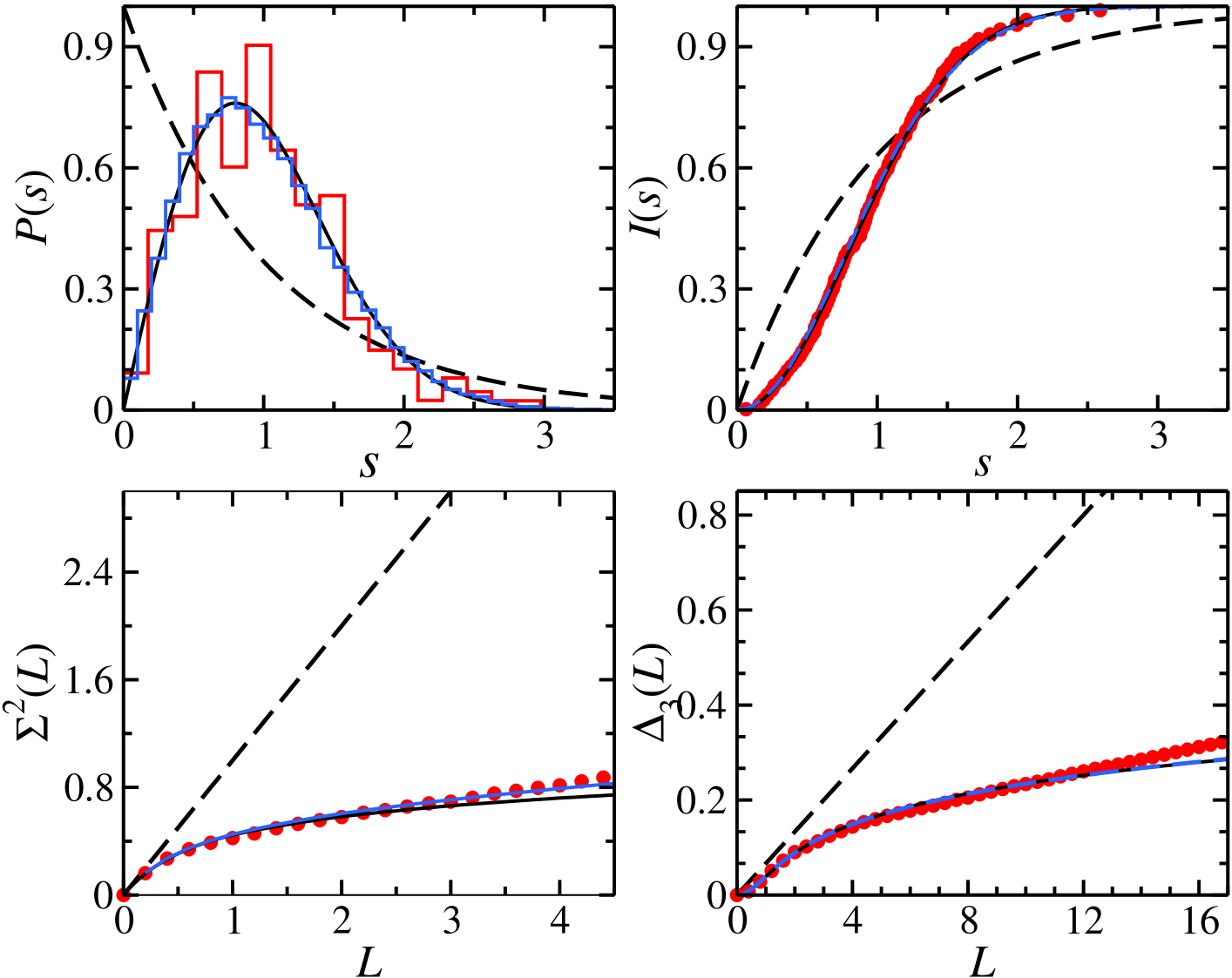}
	\caption{Spectral properties of the {\bf MB} after removing the rod, i.e., the wall, that separates it into two parts (red histogram and dots). The blue histogram and solid lines show the results for the {\bf RMT} model~\refeq{RMTModel} with coupling strengths $\lambda =3.5$.}
\label{No_Wall}
\end{figure}
We identified in the frequency range 8-12 GHz approximately $490$ resonance frequencies for three {\bf MB} and analyzed the spectral properties for resonance frequencies above 8.7~GHz. Below $f\approx 8.5$~GHz the coupling is very small (cf.~\reffig{Wavefunctions}) and then increases with frequency until it saturates at $f\gtrsim 8.7$~GHz. In a further analysis we restricted to the frequency range where the coupling can be assumed to be approximately frequency independent, $f\lesssim 11.0$~GHz, shown in blue in~\reffig{Nfluc} in the appendix. To identify that region we computed the average of the ratios $\tilde r_j$ in a sliding frequency window, shown in~\reffig{AvRatios} in the appendix, leaving approximately $340$ resonance frequencies. We analyzed the spectral properties in that frequency range, also in smaller ones and using all identified resonance frequencies above 8.7~GHz, and found up to given accuracy same results for the coupling parameter $\lambda$.   
\subsection{Comparison with the {\bf RMT} model~\refeq{RMTModel}}
In Figs.~\ref{NND}-\ref{Power} are shown the experimental results for the different statistical measures for the spectral properties as red histograms or dots. They are compared to the corresponding curves deduced from the {\bf RMT} model~\refeq{RMTModel} (blue). The parameter $\lambda$ provides a measure for the strength of coupling of the eigenmodes of the two parts of ther cavities. It was determined by comparing the experimental results for the nearest-neighbor spacing distribution $P(s)$ and the number variance $\Sigma^2(L)$ to those obtained from the {\bf RMT} model~\refeq{RMTModel}. For this we roughly estimated the value of $\lambda$ for the three cases and then varied it in steps of 0.01 around that value and analyzed the mean-square deviation for $\Sigma^2(L)$ to identify the best-fitting value of $\lambda$. Then, using that value, we checked agreement for $P(s)$ and for $\Delta_3(L)$. Here, we chose an ensemble of 200 random matrices with dimension $N_1=N_2=250$ for of $H_1$ and $H_2$, so that that of $H$ was similar to the number of resonance frequencies taken into account in the {\bf MB}s. Here, we disregarded the lowest and largest $10\%$ of the eigenvalues of $H$ and unfolded them by employing the smooth part of the integrated spectral density, which was determined by using a polynomial of fifth order for the fit. The procedure applied in~\cite{Dietz2014} does not work in the present case because the coupling is too strong. This yielded $\lambda=0.03, 0.23, 0.35$ for the {\bf MB1}, {\bf MB2}, and {\bf MB3}, respectively. 

Agreement between the experimental and {\bf RMT} results is very good for the short-range correlations and for $\Delta_3(L)$ and $R_2(r)$. Especially the number variance $\Sigma^2(L)$ is sensitive to small deviations from {\bf RMT} behavior, resulting, e.g. from non-universal features~\cite{Berry1985} associated with the shortest periodic or due to scattering at the tip of the rod which would affect the coupling~\cite{Bogomolny2006}. Discrepancies between the experimental and RMT results are observed for interval lengths beyond about $L=2-3$ mean spacings, however, not in the $\Delta_3(L)$ statistics which may be deduced from the number variance~\cite{Mehta2004}. Deviations of $K(t)$ and $s(\tau)$ from the {\bf RMT} results is attributed to the fact, that the lengths of the eigenvalue sequences are comparably short and above all, only one instead of an ensemble of eigenvalue spectra are available. Yet, it can be clearly seen that the experimental results align well with those for the {\bf RMT} model~\refeq{RMTModel}, implying that the latter may serve as a model to determine the strength of the coupling between the modes excited in the two parts of the {\bf MB}s. We also checked that, when removing the wall, the spectral properties are close to GOE statistics. The spectral properties are compared in~\reffig{No_Wall} to the corresponding curves deduced from the {\bf RMT} model~\refeq{RMTModel} with $\lambda=3.5$. For the empty circular {\bf MB} the spectral properties agree well with those of the analytic eigenvalues of the circular {\bf QB}, and accordingly for short range correlations with Poisson statistics whereas deviations are observed for long-range correlations as expected if only several hundreds or thousands of eigenvalues are taken into account~\cite{Yupei2022,Zhang2024}; cf.~\reffig{SpectraSC_Comsol} in the appendix.

\subsection{Comparison with other models for quantum systems with mixed regular-chaotic classical dynamics} 
\begin{figure}[!htbp]
\centering
\includegraphics[width=0.9\linewidth]{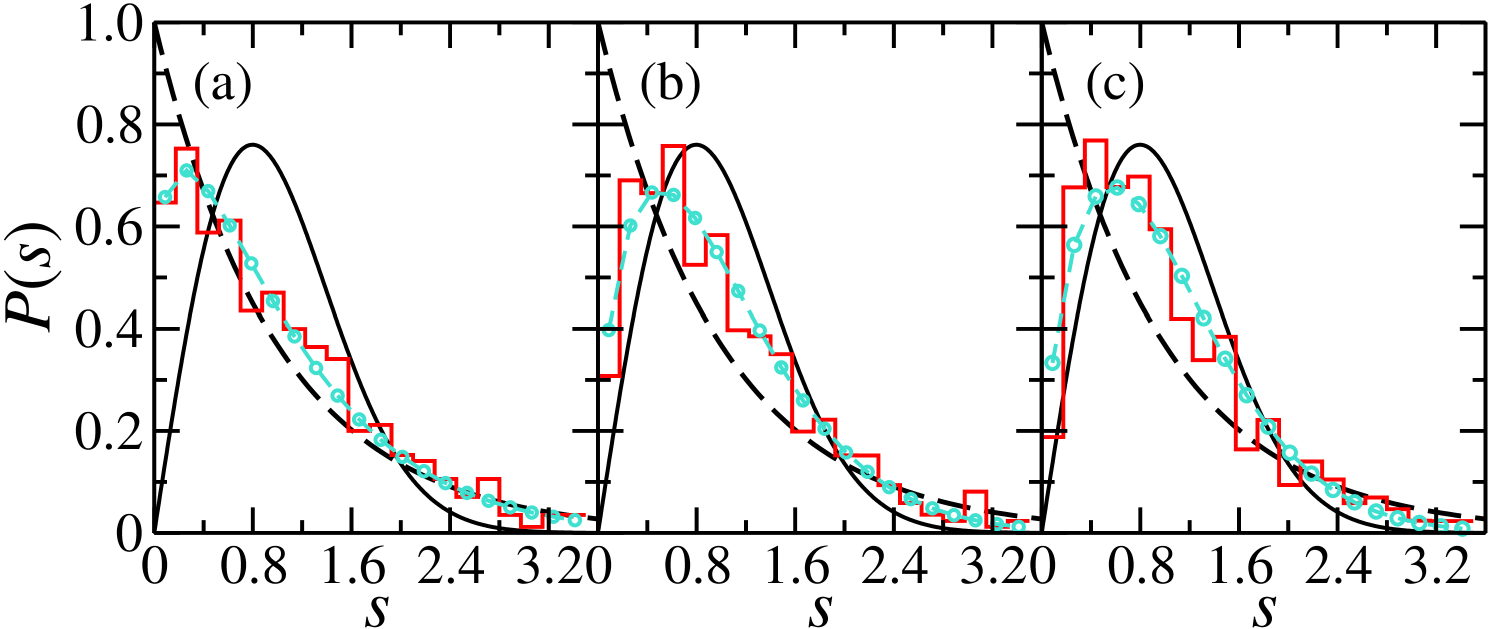}
\caption{Nearest-neighbor spacing distribution $P(s)$ for weak (a), moderate (b) and strong (c) coupling. The red histograms were obtained from the experimental data, the turquoise ones from a fit of the Brody distribution to the experimental curves, yielding for the Brody parameter $\omega =0.195063$, $\omega =0.465651$, and $\omega=0.556615$, respectively. The curves are compared with the nearest-neighbor spacing distributions for Poisson random numbers (black dashed lines) and random matrices from the GOE (black solid lines).}
\label{NND_Brody}
\end{figure}
\begin{figure}[!htbp]
\centering
\includegraphics[width=0.9\linewidth]{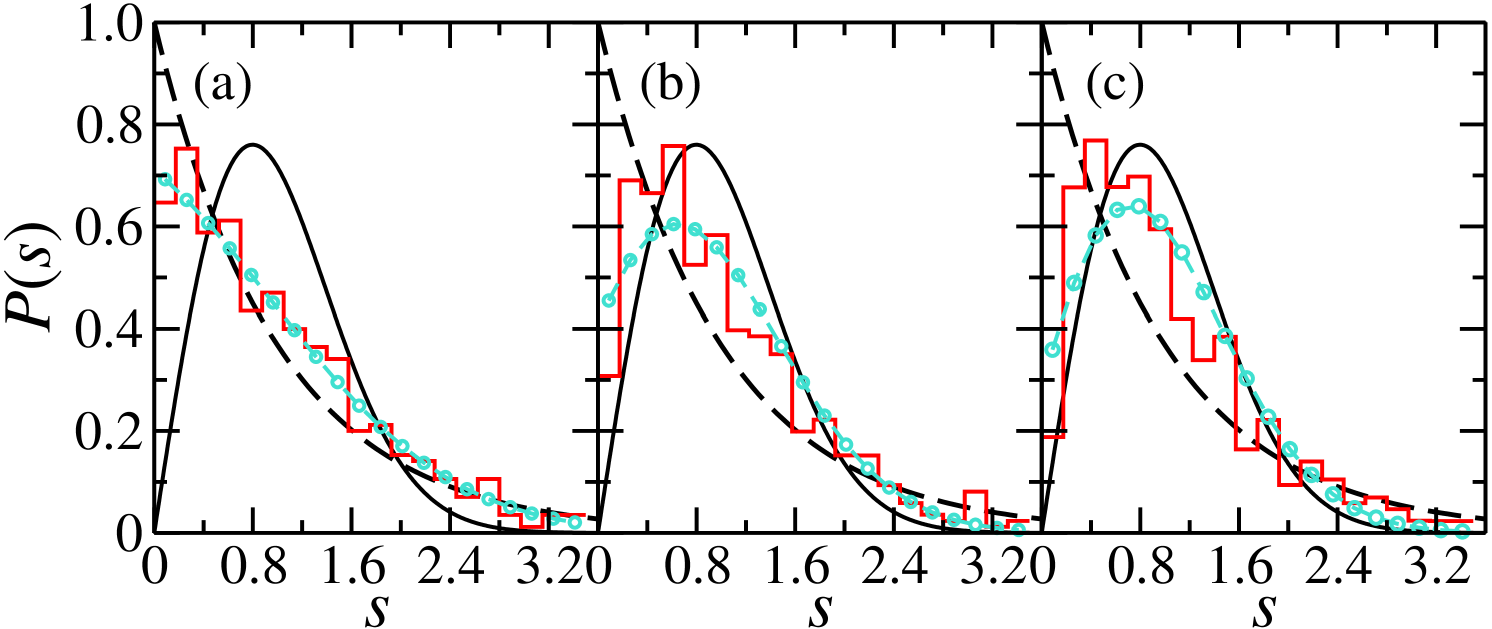}
\caption{Nearest-neighbor spacing distribution $P(s)$ for weak (a), moderate (b) and strong (c) coupling. The red histograms were obtained from the experimental data, the turquoise ones from a fit of the Berry-Robnik distribution to the experimental curves, yielding for the Berry-Robnik parameter $\mu =0.537817$, $\mu =0.770701$, and $\mu=0.846023$, respectively. The curves are compared with the nearest-neighbor spacing distributions for Poisson random numbers (black dashed lines) and random matrices from the GOE (black solid lines).}
\label{NND_BR}
\end{figure}
\begin{figure}[!htbp]
\centering
\includegraphics[width=0.9\linewidth]{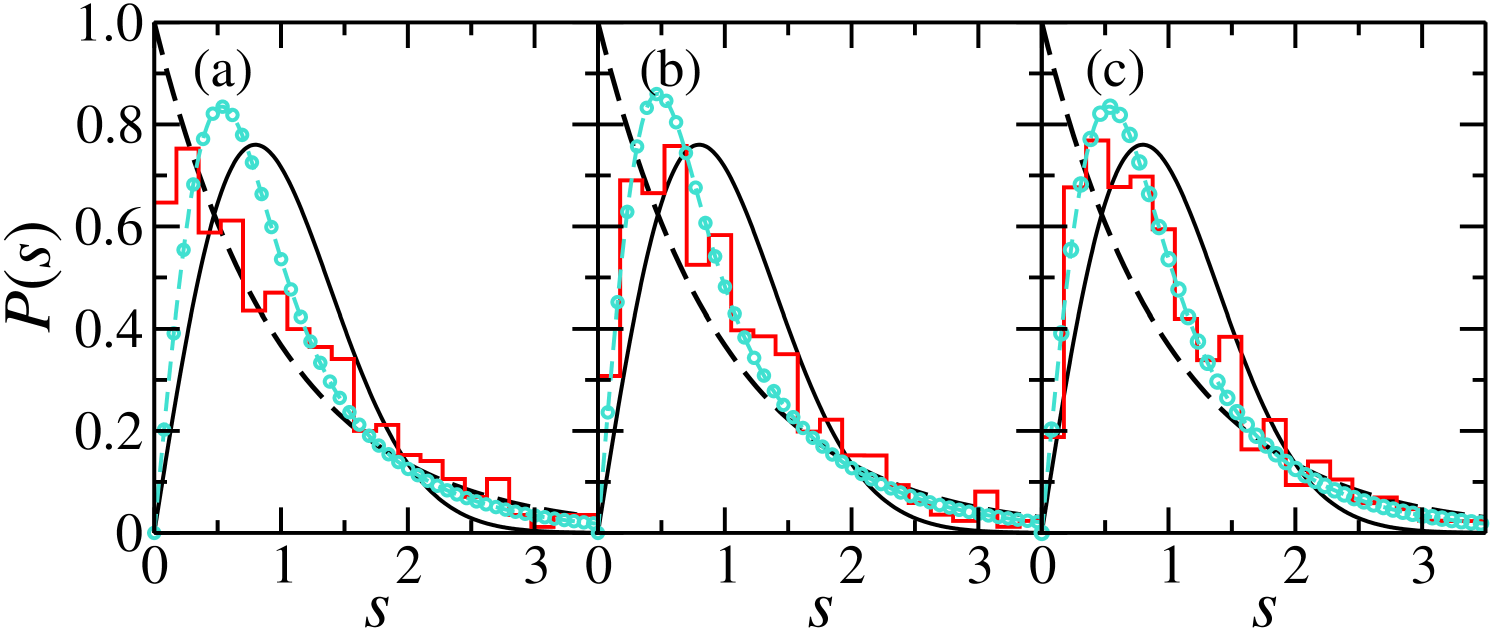}
\caption{Nearest-neighbor spacing distribution $P(s)$ for weak (a), moderate (b) and strong (c) coupling. The red histograms were obtained from the experimental data, the turquoise ones from a fit of the Lenz distribution to the experimental curves, yielding for the Lenz parameter $\lambda_L =0.09$, $\lambda_L =0.21$, and $\lambda_L=0.26$, respectively. The curves are compared with the nearest-neighbor spacing distributions for Poisson random numbers (black dashed lines) and random matrices from the GOE (black solid lines).}
\label{NND_Lenz}
\end{figure}
The Brody distribution provides an empirical description for the nearest-neighbor spacing distribution of quantum systems, whose corresponding classical dynamics comprises integrable and chaotic components. It interpolates between the Poisson distribution and Wigner surmise by varying one parameter $\omega$ between zero and unity,  
\begin{equation}
P_{\omega}(s)=(1+\omega)As^\omega\exp\left(-As^{\omega +1}\right),\, A=\left[\Gamma\left(\frac{2+\omega}{1+\omega}\right)\right]^{1+\omega}.\label{Brody} 
\end{equation}
Generally, it provides a suitable measure to estimate the degree of chaoticity. The Brody parameters obtained from a fit of this distribution to the experimental results for the nearest-neighbor spacing distributions yields the values provided in the caption of~\reffig{NND_Brody}, showing the resulting distributions together with the experimental ones, which agree quite well. 

Similarly, the Berry-Robnik distribution is applicable to quantum systems of which the phase space of the corresponding classical dynamics can be well separated into regular and chaotic regions. A prominent example with this feature are the mushroom billiards~\cite{Bunimovich2001} for which among other aspects agreement with the Berry-Robnik distribution has been checked in experiments similar to the present one with superconducting microwave billiards~\cite{Dietz2007b}. It depends on a parameter $\mu$, which gives the fraction of the energy shell, for which the classical dynamics is chaotic,
\bea
&&P_{\mu}(s)=\label{BR}
e^{-(1-\mu)s}(1-\mu)^2\Phi\left(\frac{\sqrt{\pi}}{2}\mu s\right)\\
&&+\left[2(1-\mu)\mu+\frac{\pi}{2}\mu^3s\right]e^{-\pi/4(\mu s)^2-(1-\mu)s},
\nonumber
\eea
where $\Phi(x)$ denotes the complementary error function. The values of $\mu$ obtained from the fit of this distribution to the experimental nearest-neighbor spacing distributions are provided in the captions of~\reffig{NND_BR}, where the resulting Berry-Robnik distributions are compared to the experimental ones. For non-zero coupling clear deviations are observed. This may be attributed to the fact, that in the present case the slit positions are chosen such that essentially all eigenstates of the integrable and chaotic parts have a nonvanishing overlap, implying that they even approximately cannot be separated into integrable and chaotic states~\cite{Dietz2014}.

The Lenz distribution~\reffig{NND_Lenz} has been derived based on $2\times 2$-dimensional random matrices from the Rosenzweig-Porter model for the transition from Poisson to GOE~\cite{Rosenzweig1960,Lenz1991,Cadez2024} yielding
\be
\label{PSGOE}
P^{0\to 1}(s;\tilde D)=2\tilde D^2se^{-(\tilde Ds)^2}\int_0^\infty dp e^{-\frac{p^2}{8\lambda_L^2}-p}I_0\left(\frac{\tilde Dsp}{\sqrt{2}\lambda_L}\right),
\ee
where $\tilde D=D/(2\sqrt{2}\lambda_L)$ with $D=\int_0^\infty dssP^{0\to 1}(s;\tilde D=1/[2\sqrt{2}\lambda_L])$, and $I_0(x)$ is the modified Bessel function. The distribution experiences a transition from Poisson for $\lambda_L =0$ to the Wigner surmise for the GOE for $\lambda_L\to\infty$. We also compared experimental to {\bf RMT} results for other statistical measures and didn't find good agreement, implying that this Rosenzweig-Porter model is not applicable to the present case. In~\refsec{Comsol} of the appendix we compare experimental results with numerical ones deduced from the resonance frequencies obtained with COMSOL Multiphysics. 

\section{Conclusions\label{Concl}}
We performed experiments with superconducting {\bf MB}s modeling coupled {\bf QB}s with integrable and chaotic dynamics, respectively. Coupling is realized by introducing an opening of varying width into a common wall, so that electric-field modes, that is wave functions, can penetrate from one billiard into the other one. We consider three cases denoted as very weak, moderate and strong coupling. In the case of weakest coupling the eigenmodes of the two parts of {\bf MB}1 are coupled due to the leakage resulting from the non-perfect adjustment of the separating wall and the top and bottom plates. For that situation the eigenvalue spectrum essentially corresponds to a superposition of those of the two cavities, as indicated by the {\bf RMT} model. On the other hand, even for the case with strongest coupling the spectral properties do not agree with GOE statistics and are closer to Poisson statistics, implying that the coupling still is comparably weak. These findings demonstrate that already a weak mixing of the eigenmodes of the integrable and chaotic parts leads to deviations from the statistics for completely uncoupled cavities. Essentially, coupling of these eigenmodes leads to a breaking of the symmetry present in the billiard with integrable classical dynamics, and an intrusion of regular components into the chaotic one. We demonstrate that the spectral properties are well described by random matrices from a variant of the Rosenzweig-Porter model~\cite{Rosenzweig1960}, whose structure is similar to those employed for symmetry breaking in~\cite{Dietz2006a}. They consist of two diagonal blocks containing Hermitean matrices with Poisson and GOE statistics, respectively, that are coupled via off-diagonal block matrices. The coupling strength is tuned by a parameter $\lambda$. Thus we may use this {\bf RMT} model to determine the coupling strength of the eigenmodes of the two parts of the {\bf QB}, or equivalently, {\bf MB}. Our findings indicate, that generally, this {\bf RMT} model is suitable to describe the effect of integrable components on a quantum system with chaotic dynamics when their wave functions overlap~\cite{Dietz2006a,Aberg2008}.    

\section{Acknowledgments}
This work was supported by the NSF of China under Grants No. 11775100, No. 12247101, and No. 11961131009. X.Z. acknowledges financial support from the China Scholarship Council (Grant No. CSC-202306180087). B.D. and X.Z. acknowledge financial support from the Institute for Basic Science in Korea through Project No. IBS-R024-D1.
\bibliography{References,Chaos-Refs,manual_refs}

\begin{appendix}
\renewcommand{\theequation}{A\arabic{equation}}
\renewcommand{\thefigure}{A\arabic{figure}}
\setcounter{figure}{0}
\section{Details of geometry of the microwave billiard\label{Details}}
\begin{figure}[!htbp]
\centering
\includegraphics[width=0.9\linewidth]{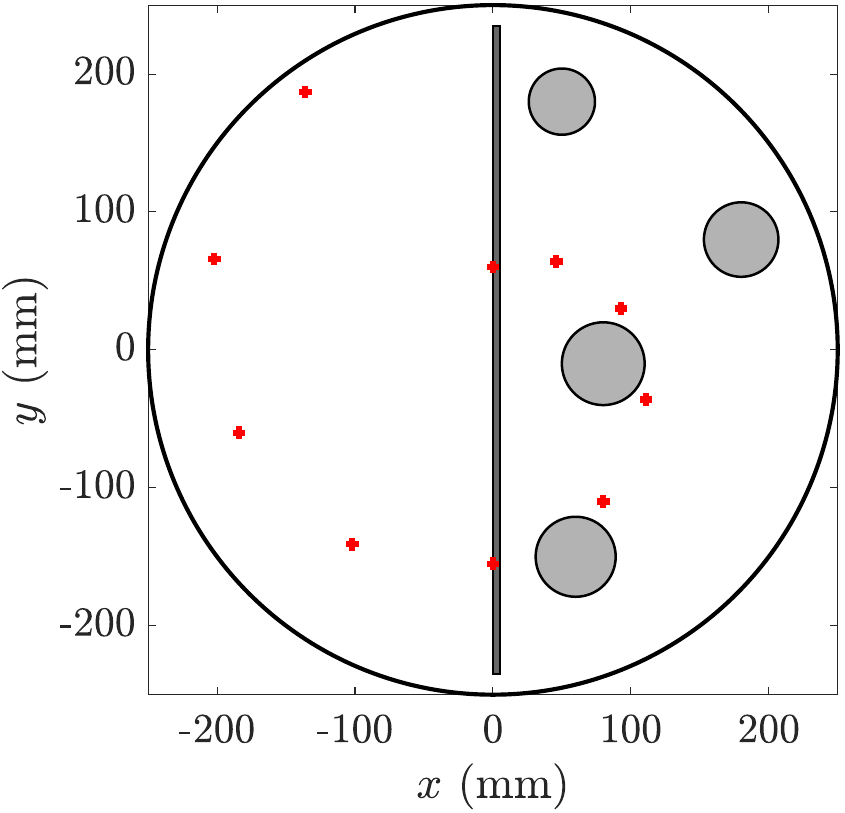}
\caption{Schematic of the microwave cavity.}
\label{Model}
\end{figure}
The microwave cavity has the shape of a circle with diameter $D = 500~\mm$ (radius $R = 250~\mm$) and a height $h = 5~\mm$. It is separated into two parts by a rectangular metallic rod of varying length, which is positioned with one wall along a diameter such that it leaves two openings of equal width between its ends and the circular wall, thus facilitating a coupling between the eigendmodes in the two parts. The height of the rod equals that of the cavity, also the width $w = 5~\mm$. We consider three lengths $L$ of the rod,
\begin{itemize}
  \item weak coupling: $L = 500~\mm$,
  \item moderate coupling: $L = 476~\mm$,
  \item strong coupling: $L = 470~\mm$.
\end{itemize}
The rod is squeezed between the top and bottom plates of the cavity, but is not fit perfectly flush with them, so that there is microwave leakage from one part of the cavity to the other one even in the case for $L=500$~mm. Four cylindrical scatterers are inserted in the nearly semicircular part of the cavity. Their centers are located at the following positions with respect to the cavity center,
\[
(180,\,80)~\mm, (60,\,-150)~\mm, (80,\,-10)~\mm, (50,\,180)~\mm.
\]
The corresponding scatterer radii are
\[
r_1 = 27~\mm, r_2 = 29~\mm, r_3 = 30~\mm, r_4 = 24~\mm.
\]

All antenna positions are defined in a Cartesian coordinate system with the cavity center located at the origin $(0,0)$. The holes for the antennas shown in Fig.~1 (left panel) are listed below, where the first column denotes the $x$ coordinate and the second column
denotes the $y$ coordinate (units in mm):
\begin{equation}
\begin{aligned}
\text{Antenna 1:} & \quad ( 0, 60 ), \\
\text{Antenna 2:} & \quad (  46, 64 ), \\
\text{Antenna 3:} & \quad (  93, 30 ), \\
\text{Antenna 4:} & \quad ( 111, -36 ), \\
\text{Antenna 5:} & \quad (  80, -110 ), \\
\text{Antenna 6:} & \quad (  0, -155 ), \\
\text{Antenna 7:} & \quad ( -102, -141 ), \\
\text{Antenna 8:} & \quad ( -184, -60 ), \\
\text{Antenna 9:} & \quad ( -202, 66 ), \\
\text{Antenna 10:} & \quad ( -136, 187).
\end{aligned}
\end{equation}
The holes for antennas 1 and 6 are in the region of the rod, and thus they are not used.

The area $\mathcal{A}$ of the cavity is given by the area of the circular
tud minus the areas occupied by the scatterers and by the metallic rod
\begin{equation}
\mathcal{A}(L)=\mathcal{A}_{\mathrm{circle}}-\mathcal{A}_{\mathrm{scat}}-\mathcal{A}_{\mathrm{bar}}(L),
\end{equation}
with 
\begin{equation}
\mathcal{A}_{\mathrm{circle}}=\pi R^2
=1963.4954~\cm^2, 
\end{equation}
\begin{equation}
\mathcal{A}_{\mathrm{scat}}=\sum_{i=1}^{4}\pi r_i^2
=95.6929~\cm^2, 
\end{equation}
and,
\begin{equation}
\mathcal{A}_{\mathrm{bar}}(L)=wL.
\end{equation}
With $w=5~\mm$ we obtain
\begin{align}
\mathcal{A}(500~\mm) 
=1842.8025~\cm^2,\\
\mathcal{A}(476~\mm) 
=1844.0025~\cm^2,\\
\mathcal{A}(470~\mm) 
=1844.3025~\cm^2.
\end{align}

\section{More experimental results\label{MExp}}
\begin{figure}[htbp]
\includegraphics[width=0.9\linewidth]{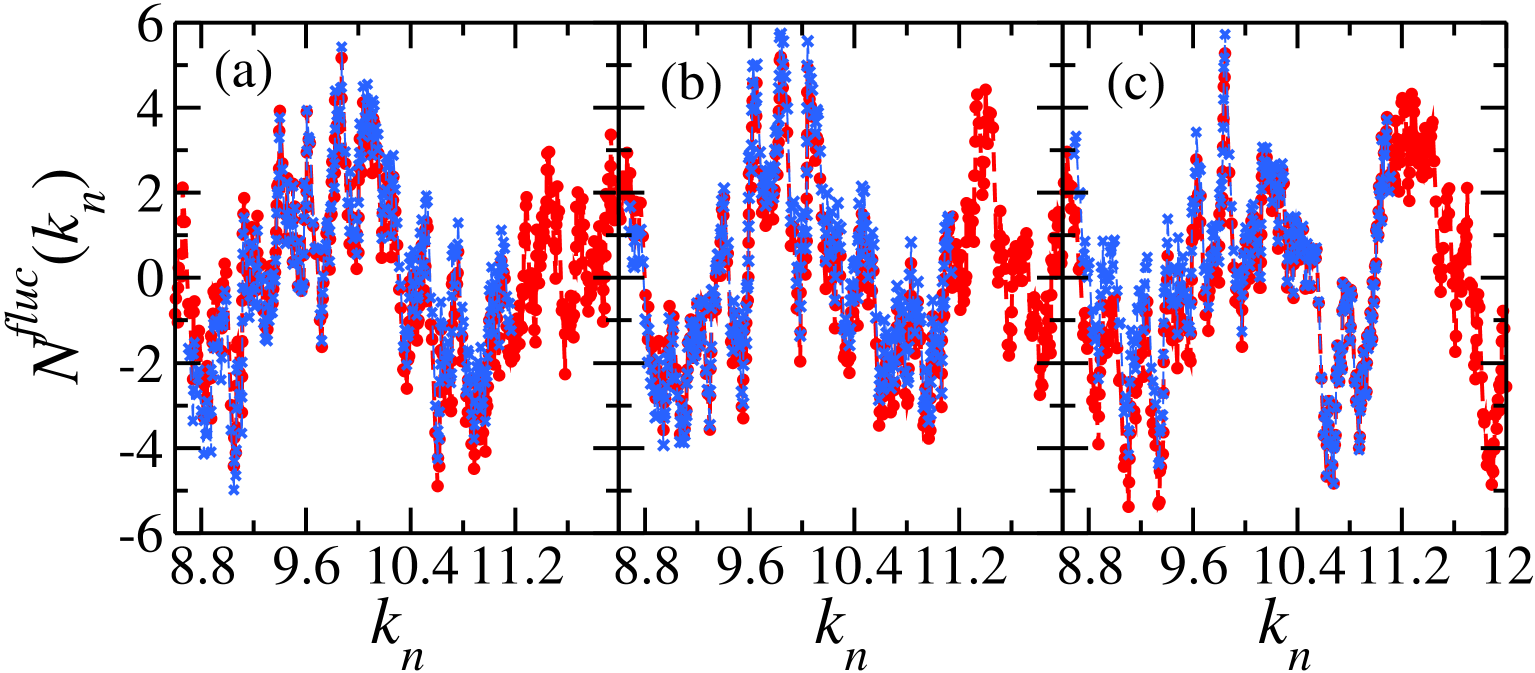}
	\caption{The fluctuating part of the integrated spectral density for weak (a), moderate (b) and strong (c) coupling. The red dots show the results obtained when considering all resonance frequencies, the blue crosses when restricting to the range of approximately constant average ratios $<\tilde r_n>$, shown in~\reffig{AvRatios}.}
\label{Nfluc}
\end{figure}

\begin{figure}[htbp]
\includegraphics[width=0.7\linewidth]{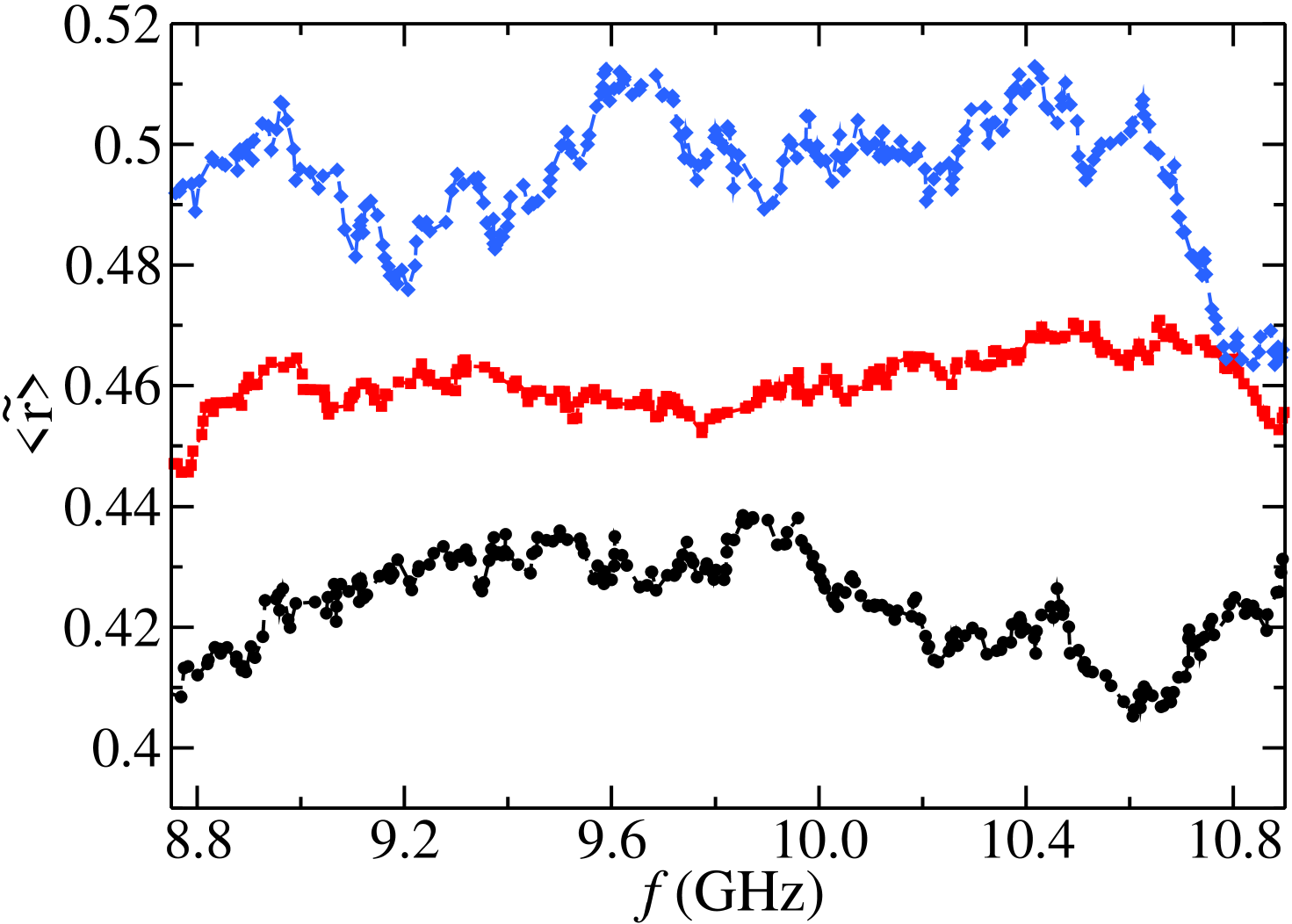}
	\caption{Average ratios $<\tilde r_n>$ for weak (black), moderate (red) and strong (blue) coupling. They were determined in a sliding frequency window containing 150 resonance frequencies.}
\label{AvRatios}
\end{figure}

In this section we exhibit the fluctuating part of the integrated spectral density~\reffig{Nfluc}, resulting from an unfolding of all resonance frequencies identified in the range $[8.7,12]$~GHz and those in the frequency range $[8.7,11]$~GHz where the coupling strength can be assumed to be frequency independent. To find that region we computed the average ratios $<\tilde r_n>$ in a sliding frequency interval containing 150 resonance frequencies; cf.~\reffig{AvRatios}. Furthermore, to test the frequency dependence of the parameter $\lambda$, which gives a measure for the coupling, we analyzed the spectral properties above 8.7~GHz in intervals containing 200, 300 and all identified resonance frequencies, and obtained in all cases the same result up to a few percent of the value.
\section{Comparison of experimental results with Comsol Computations\label{Comsol}}
\begin{figure}
\includegraphics[width=0.9\linewidth]{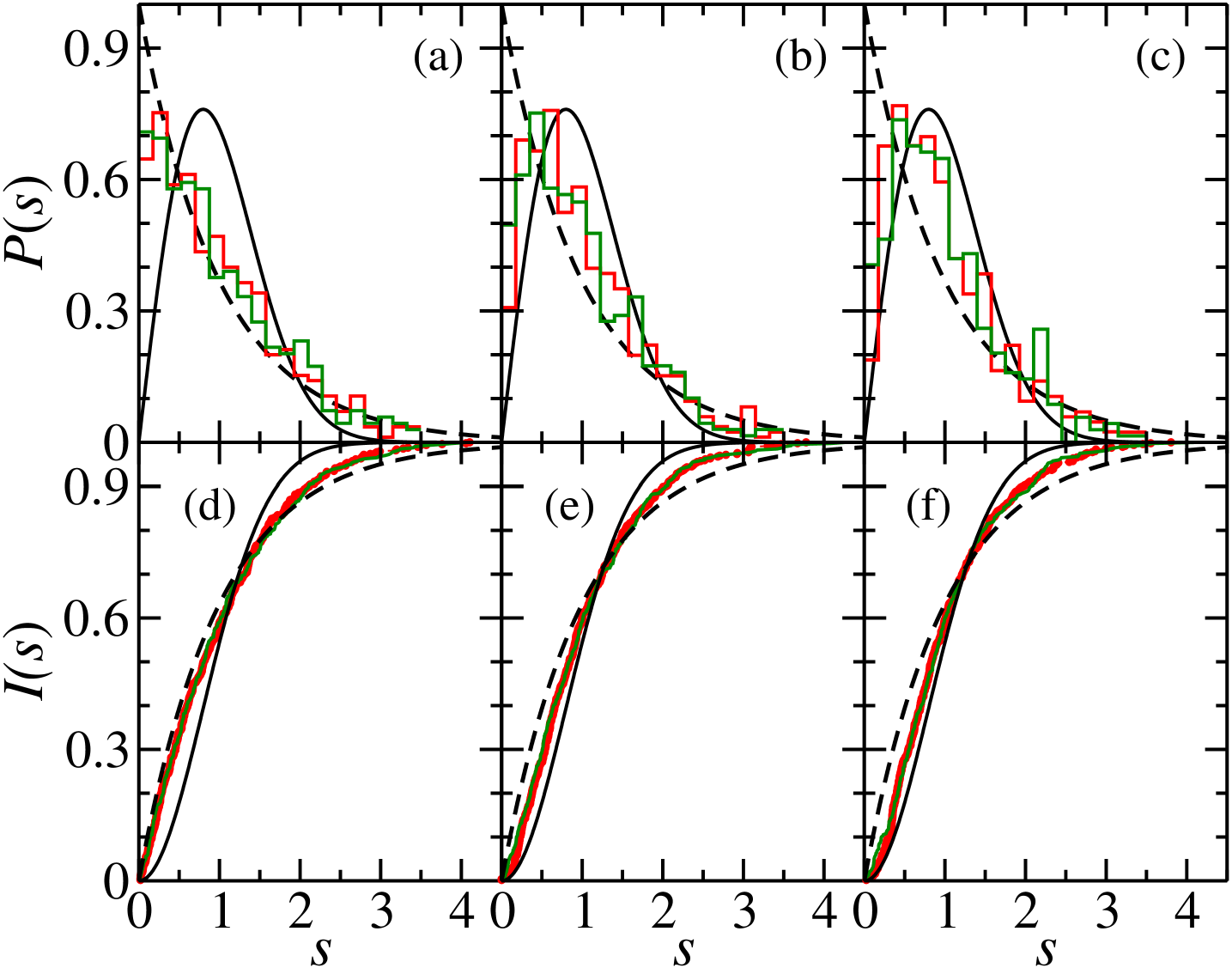}
\caption{Nearest-neighbor spacing distribution $P(s)$ [(a)-(c)] and the corresponding cumulative distribution $I(s)$ [(d)-(f)] for weak [(a),(d)], moderate [(b),(e)] and strong [(c),(f)] coupling. The red histograms and dots were obtained from the experimental data, the green ones from COMSOL Multiphysics computations, respectively. These are compared with corresponding distributions for Poisson random numbers (black dashed lines) and random matrices from the GOE (black solid lines).}
\label{NND_Comsol}
\end{figure}

\begin{figure}[!htbp]
\centering
\includegraphics[width=0.9\linewidth]{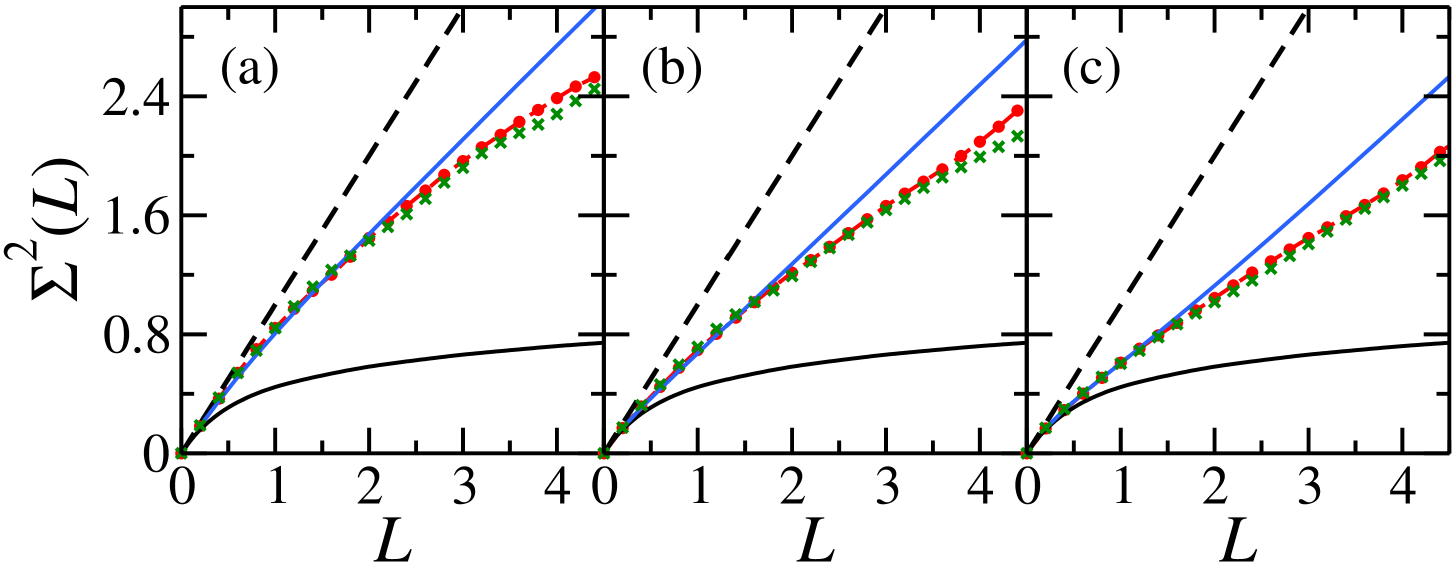}
	\caption{Same as~\reffig{NND_Comsol} for the number variance $\Sigma^2(L)$. The blue ones show the corresponding curves for the {\bf RMT} model~\refeq{RMTModel} for coupling strengths (a) $\lambda =0.03$, (b) $\lambda =0.23$, and (c) $\lambda=0.35$}
\label{Sigma2_Comsol}
\end{figure}

\begin{figure}[!htbp]
\centering
\includegraphics[width=0.9\linewidth]{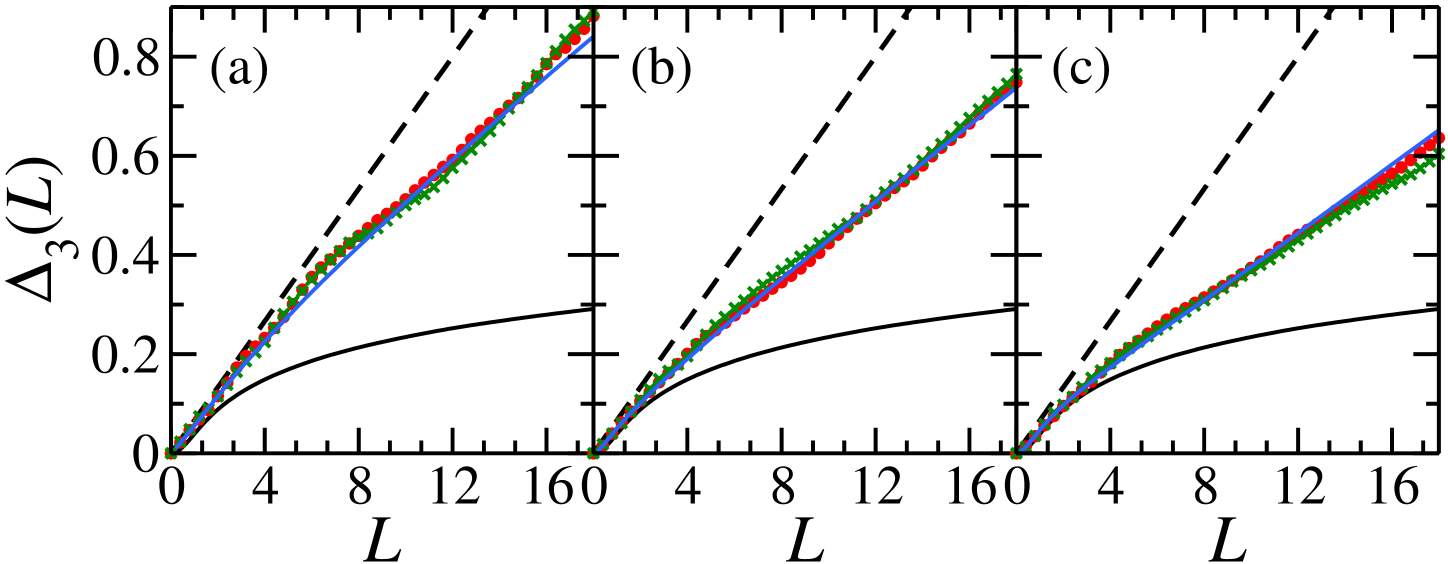}
\caption{Same as~\reffig{Sigma2_Comsol} for the rigidity $\Delta_3(L)$.}
\label{Delta3_Comsol}
\end{figure}

Besides the electric-field distributions COMSOL Multiphysics provides the resonance frequencies. We analyzed the spectral properties and compared them to the experimental ones; cf. Figs.~\ref{NND_Comsol},~\ref{Sigma2_Comsol} and~\ref{Delta3_Comsol}. The numerical and experimental results agree very well with each other.  

\begin{figure}[!htbp]
\centering
\includegraphics[width=0.9\linewidth]{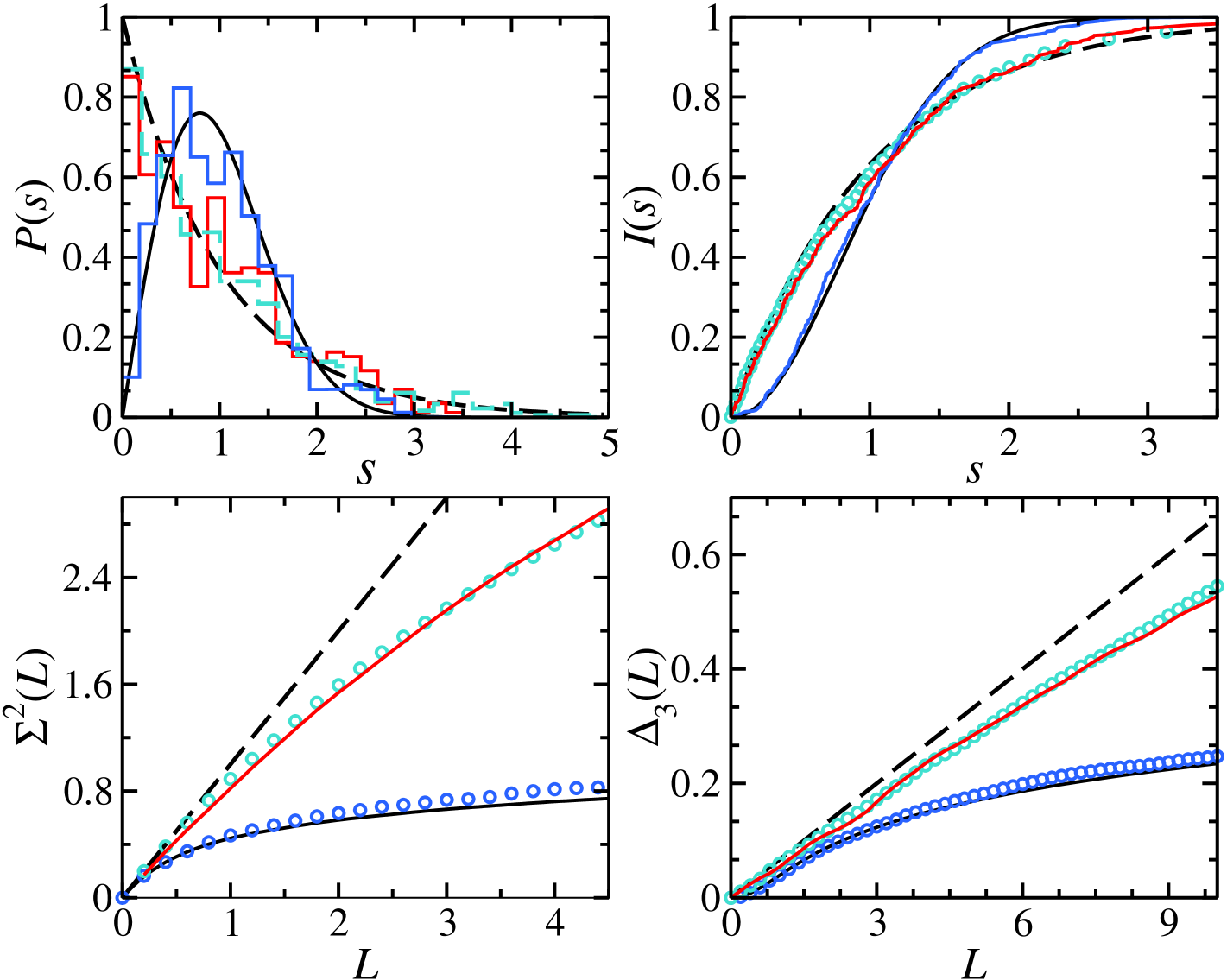}
	\caption{Spectral properties of the semicircular part (red histogram and circles) and the chaotic part (blue histograms and circles). The former are compared to the spectral properties obtained from the analytical eigenvalues.}
\label{SpectraSC_Comsol}
\end{figure}

As mentioned in the main text, even for the case, where the rod length coincides with the diameter length a very weak coupling is present. Therefore, we were not able to separate eigenvalues for that case into those of the semicircular and chaotic parts. To check agreement with Poissonian statistics and GOE statistics, respectively, we used data obtained from COMSOL Multiphysics. For the semicircular part we compared the results with those of the analytically computed eigenvalues. Note, that in Refs.~\cite{Zhang2023b} we performed measurements with the empty circular cavity and compared the spectral properties with those obtained from the analytically determined eigenvalues and found excellent agreement. Similarly, in the present case we find very good agreement as demonstrated in~\reffig{SpectraSC_Comsol}. In both cases agreement of the long-range correlations with those of Poissonian random numbers is as good as expected for the comparatively small number of resonance frequencies.  
\end{appendix}

\end{document}